\documentclass[acmsmall,screen,nonacm]{acmart}

\usepackage{booktabs}
\usepackage{subcaption}
\usepackage{listings}
\usepackage{xcolor}
\usepackage{tcolorbox}
\usepackage{microtype}
\usepackage{llvm/lang} 
\usepackage{nasm/style} 
\usepackage{xcolor,colortbl}
\usepackage{caption}
\usepackage{courier}
\usepackage{multicol}
\definecolor{codebg}{rgb}{0.95,0.95,0.95}
\definecolor{bluekeywords}{rgb}{0.13,0.13,1}
\definecolor{candypink}{rgb}{0.89, 0.44, 0.48} 
\lstdefinestyle{mlir}{
    basicstyle=\ttfamily\footnotesize,
    language=llvm, style=nasm,
    numbers=left, 
    stepnumber=1, 
    tabsize=1,
    backgroundcolor=\color{codebg},
    numberstyle=\tiny,
    frame=single,
    framesep=1pt,
    rulecolor=\color{black},
    breaklines=true,
    breakatwhitespace=true,
    escapeinside={*@}{@*},
    commentstyle=\color{gray!80},
}

\lstdefinestyle{wasm}{
    basicstyle=\ttfamily\footnotesize,
    numbers=left, 
    stepnumber=1, 
    tabsize=1,
    backgroundcolor=\color{codebg},
    numberstyle=\tiny,
    frame=single,
    framesep=1pt,
    rulecolor=\color{black},
    breaklines=true,
    breakatwhitespace=true,
    escapeinside={*@}{@*},
    comment=[l]{;;},
    commentstyle=\color{gray!80},
    otherkeywords={},
    morekeywords=[1]{i32,f32,i64,f64},
    keywordstyle={[1]\color{violet}},
    morekeywords=[2]{0},
    keywordstyle={[2]\color{violet}},
    morekeywords=[3]{add,const}
    keywordstyle={[3]\color{bluemunsell}},
    morekeywords=[4]{local},
    keywordstyle={[4]\color{candypink}},
    morekeywords=[5]{module, func, param, result, global, get_global, mut, set_global, export, import, memory, data, get_local, set_local, elem, table, call,call_indirect, type, block, loop, br, br_if, br_table, return, if, else, end, then, else, end },
    keywordstyle={[5]\color{bluekeywords}},
    morekeywords=[6]{=,;},
    keywordstyle={[6]\color{britishracinggreen}},
    morekeywords=[7]{(,),[,],.},
    keywordstyle={[7]\color{black}},
}
\lstdefinestyle{other}{
    basicstyle=\ttfamily\footnotesize,
    numbers=left, 
    stepnumber=1, 
    tabsize=1,
    backgroundcolor=\color{codebg},
    numberstyle=\tiny,
    frame=single,
    breaklines=true,
    breakatwhitespace=true,
    escapeinside={*@}{@*},
	comment=[l]{;;},
    commentstyle=\color{gray!80},
}
\usepackage{courier}

\usepackage{xspace}
\usepackage{multirow}     
\usepackage{siunitx}
\usepackage{colortbl}
\usepackage[inline]{enumitem}
\usepackage{array}

\newif \ifnotes
\notesfalse
\ifnotes
\usepackage{todonotes}
\else \usepackage[disable]{todonotes}
\fi

\newtcolorbox{summarybox}{
    boxrule = 1.5pt,
    colframe = black 
}

\newcommand{\cellyes}{\cellcolor{green!25}{$\checkmark$}}
\newcommand{\cellno}{\cellcolor{red!25}{$\times$}}

\newcommand{\cellna}{N/A}

\newcommand{\styledialect}[1]{\texttt{#1}\xspace}

\newcommand{\ssawasm}{\styledialect{SsaWasm}}
\newcommand{\wasm}{\styledialect{Wasm}}
\newcommand{\tosa}{\styledialect{TOSA}}
\newcommand{\arith}{\styledialect{Arith}}
\newcommand{\linalg}{\styledialect{LinAlg}}
\newcommand{\memref}{\styledialect{MemRef}}
\newcommand{\func}{\styledialect{Func}}
\newcommand{\scf}{\styledialect{SCF}}
\newcommand{\cf}{\styledialect{CF}}
\newcommand{\llvm}{\styledialect{LLVM}}
\newcommand{\dcont}{\styledialect{DCont}}

\newcommand{\async}{\styledialect{Async}}

\newcommand{\tool}{WAMI\xspace}
\newcommand{\paragraphb}[1]{\vspace{5pt}\noindent{\bf #1}}

\author{Byeongjee Kang}
\affiliation{
    \institution{Carnegie Mellon University}
    \city{Pittsburgh}
    \country{USA}
}
\email{byeongjee@cmu.edu}
\author{Harsh Desai}
\affiliation{
    \institution{Yale University}
    \city{New Haven}
    \country{USA}
}
\email{harsh.desai@yale.edu}
\author{Limin Jia}
\affiliation{
    \institution{Carnegie Mellon University}
    \city{Pittsburgh}
    \country{USA}
}
\email{liminjia@andrew.cmu.edu}
\author{Brandon Lucia}
\affiliation{
    \institution{Carnegie Mellon University}
    \city{Pittsburgh}
    \country{USA}
}
\email{blucia@andrew.cmu.edu}

\title{WAMI: Compilation to WebAssembly through MLIR without Losing Abstraction}

\begin{document}

\begin{abstract}

WebAssembly (Wasm) is a portable bytecode format that serves as a compilation
target for high-level languages, enabling their secure and efficient execution
across diverse platforms, including web browsers and embedded systems.
To improve support for high-level languages without incurring significant code size
or performance overheads, Wasm continuously evolves by integrating high-level
features such as Garbage Collection and Stack Switching.
However, existing compilation approaches either lack reusable design---requiring
redundant implementation efforts for each language---%
or lose abstraction by lowering high-level constructs into low-level
shared representations like LLVM IR, which hinder the adoption of high-level
features. 
MLIR compiler infrastructure provides the compilation pipeline with multiple levels
of abstraction, preserving high-level abstractions throughout the compilation
pipeline, yet the current MLIR pipeline relies on the LLVM backend for Wasm code
generation, thereby inheriting LLVM's limitations.

This paper presents a novel compilation pipeline for Wasm, featuring
Wasm dialects explicitly designed to represent high-level Wasm
constructs within MLIR. 
Our approach enables direct generation of high-level Wasm code from
corresponding high-level MLIR dialects without losing abstraction, providing
a modular and extensible way to incorporate high-level Wasm features.
We illustrate this extensibility through a case study that leverages Stack
Switching, a recently introduced high-level feature of Wasm. 
Performance evaluations on PolyBench benchmarks show that 
our pipeline, benefiting from optimizations within the MLIR and Wasm ecosystems,
produces code with at most 7.7\% slower, and faster in some execution
environments, compared to LLVM-based compilers.

\end{abstract}

\maketitle
\section{Introduction}
\label{sec:intro}
WebAssembly (Wasm)~\cite{webassembly} is a portable bytecode format for a
virtual machine,
enabling sandboxed environments with near-native performance~\cite{haas2017bringing}.
Wasm can be deployed and executed across various platforms, including all
mainstream web browsers, desktop computers, servers, and embedded systems.
Due to these advantages,
many high-level programming languages, including C/C++, Rust, Go, and Haskell,
support Wasm as a compilation target~\cite{emscripten, haskell-ghc, go-wasm},
and the number of languages targeting Wasm is increasing~\cite{kotlinwasm,
javawasm, dartwasm, andres2023wasocaml}.

The Wasm specification continues to integrate high-level features
that lack direct counterparts in traditional assembly instructions.
This is driven by the push for improved support for high-level languages,
enhanced interoperability with JavaScript, and improved security and performance.
For instance, the Garbage Collection feature, now enabled by default across
major browsers, introduces specialized types and instructions for host-managed
heap-allocated objects, supporting compilation of garbage-collected languages to
Wasm without the overhead of bundling a runtime garbage collector.
Similarly, numerous Wasm feature proposals that incorporate high-level language constructs, such as Exception
Handling~\cite{wasm-exception-handling}, Stack Switching~\cite{wasm-stack-switching}, Threads~\cite{wasm-threads}, Multiple
Memories~\cite{wasm-multiple-memories}, and Reference-Typed Strings~\cite{wasm-reference-typed-strings}, 
are being actively proposed, discussed, and developed 
to be standardized in the Wasm specification.

However, existing compilation approaches for translating
high-level languages into Wasm lack
adequate support for quick adoption of new Wasm features. 
Existing approaches generally fall into two categories:
building a custom compilation pipeline for the language and leveraging a
compiler infrastructure, such as LLVM.
Custom compilation pipelines allow flexible design, but lack community support
that shared compiler infrastructure can provide, requiring
significant one-off implementation
efforts for each language.
On the other hand, compiler infrastructures such as LLVM~\cite{llvm} offer
a shared infrastructure that can be reused across different languages,
but have limited support for high-level Wasm features.
LLVM compiles high-level languages into an intermediate representation, LLVM IR,
shared by all compilation targets, leading to lose of abstractions that are
required to utilize high-level features in Wasm.

The MLIR (Multi-Level Intermediate Representation)
compiler infrastructure~\cite{mlir}
was introduced to address the abstraction loss problem of existing
compiler infrastructures. 
Unlike LLVM, which uses a singular IR, MLIR comprises multiple IR dialects
representing various abstraction levels. 
Compilation in MLIR involves dialect conversion passes rather than a monolithic 
compilation process, promoting
modularity and extensibility and enabling optimizations that are accessible only
at higher abstraction levels. 
Despite these advantages, however, the current MLIR-to-Wasm pipeline still
suffers from abstraction loss by relying on LLVM IR and LLVM's backend for Wasm
code generation.

To improve the compilation ecosystem for Wasm, we propose a new compilation pipeline for Wasm,
\emph{\tool} (WebAssembly Mlir Ir).
We introduce new dialects in MLIR to model Wasm within MLIR directly, lifting
Wasm from being a compilation target of LLVM to being a first-class citizen of
the MLIR ecosystem.
This strategy enables the direct generation of Wasm from MLIR dialects with
appropriate abstractions, preserving abstraction layers and providing a modular
framework to seamlessly incorporate high-level Wasm features.

We demonstrate the extensibility of \tool through an example.
We design a new dialect for expressing non-local control flow
and implement a dialect conversion pass to translate it into the Wasm dialect,
utilizing instructions proposed in the Stack Switching proposal.
Although the same functionality can be achieved without the high-level feature
by using the LLVM coroutine intrinsics and transformations, utilizing the
feature provides better performance, size, and abstraction.

We evaluate the performance of \tool on PolyBench~C~\cite{polybench}, a widely
used benchmark suite featuring diverse computational behaviors, translated to
MLIR using Polygeist~\cite{moses2021polygeist}.
Our evaluation spans two distinct hardware environments: a high-performance
MacBook Pro with Apple M2 Pro processor running Wasmtime~\cite{wasmtime}, and a
low-power microcontroller, Apollo4 Blue Plus, running Wasm Micro Runtime
(WAMR)~\cite{wamr}.
On both platforms, we conducted tests using both interpretation and
Ahead-of-Time (AoT) compilation approaches.
Our results show that \tool produces code that has reasonable performance
compared to code generated by LLVM, leveraging optimizations within the MLIR and
Wasm ecosystems.
These results primarily benefit from existing MLIR optimization passes
and Wasm binary optimizations available through the Binaryen toolchain~\cite{binaryen},
which we can leverage for free by being part of the MLIR and Wasm ecosystem.
Our results indicate that \tool is a viable direction for Wasm
compilation that can produce practical results without potentially lengthy
wait for advanced LLVM optimizations to be ported.

The contributions of this paper are summarized as follows:
\begin{enumerate}[topsep=1ex,partopsep=1ex,parsep=1ex]
    \item Development of novel MLIR dialects explicitly modeling Wasm
    (Section~\ref{sec:design}) and corresponding dialect conversion passes for
    compiling high-level dialects into Wasm (Section~\ref{sec:dialect-conversion}).

    \item A proof-of-concept MLIR dialect for non-local control flow
    with a dialect conversion pass to translate it into the Wasm dialect
    using Stack Switching (Section~\ref{sec:casestudy}).

    \item Performance evaluation with PolyBench C
    on two distinct hardware platforms, showing that 
    \tool generates code with robust performance
    compared to LLVM-based compilers,
    despite not relying on LLVM optimizations passes
    (Section~\ref{sec:evaluation}).
\end{enumerate}

\section{Background and Motivation}
\label{sec:motivation}

In this section, we discuss the motivations behind our work. 
First, we outline the unique characteristics of Wasm that
differentiate it from traditional assembly formats. 
Next, we highlight the limitations faced by existing compilers targeting Wasm,
arising from these unique characteristics. 
Finally, we introduce the MLIR compiler infrastructure, explaining how it
improves upon previous approaches yet still falls short of fully addressing the
specific challenges we aim to resolve.

\subsection{WebAssembly Features and Compilation Pipelines}

\subsubsection{Characteristics of Wasm}

\paragraph{Evolving Nature of Wasm}

Wasm's implementation as a virtual machine target enables it to evolve with new
capabilities over time, unlike traditional assembly formats that remain tightly
coupled with their underlying hardware.
Since the Wasm's Minimal Viable Product (MVP) was released in 2015, 8
feature proposals have been merged to the specification (Finished Proposals),
8 features, not yet merged, but supported by most runtimes (Phase 5), and 32
features are being discussed and developed (from Phase 1 to 4) \cite{wasmproposals}.
These proposals serve diverse purposes: some, like JavaScript BigInt to Wasm
i64 integration, improve interoperability with JavaScript in web environments;
others, such as relaxed SIMD and Branch Hinting, focus on performance
optimization; while features like Garbage Collection expand support for
high-level programming languages.

\paragraph{High-Level Features of Wasm}

Both the MVP and extensions of Wasm often come with instructions with higher
abstraction level and more semantic information than traditional assembly
formats.
For example, Wasm only allows structured control flow,
i.e., it does not allow jumps to arbitrary program locations,
for benefits such as control-flow integrity \cite{wasm-docs-security}.
The Garbage Collection feature introduces garbage-collected heap objects with
nontrivial types, such as arrays and structs, as first-class citizens in Wasm,
which allows languages with garbage collection to compile to Wasm more efficiently.
The Multiple Memories feature introduces multiple isolated memory spaces to Wasm,
which enables better memory isolation and management.
The Stack Switching feature, which is an early stage proposal, introduces
instructions to save and restore the execution context, so that non-local
control flow in high-level languages (e.g., coroutines) can be supported efficiently.

\subsubsection{Existing Compilers that Target Wasm}

Most languages support compilation to Wasm by either
\begin{enumerate*}[label=(\arabic*)]
    \item building a custom compiler, or
    \item utilizing LLVM compiler infrastructure.
\end{enumerate*}
Each of these approaches has its limitations in terms of supporting Wasm's frequently updated features or taking advantage of its higher-level abstractions than traditional assembly formats.
\begin{figure}[h]
    \centering
    \begin{subfigure}[b]{0.48\textwidth}
        \centering
        \includegraphics[width=\textwidth]{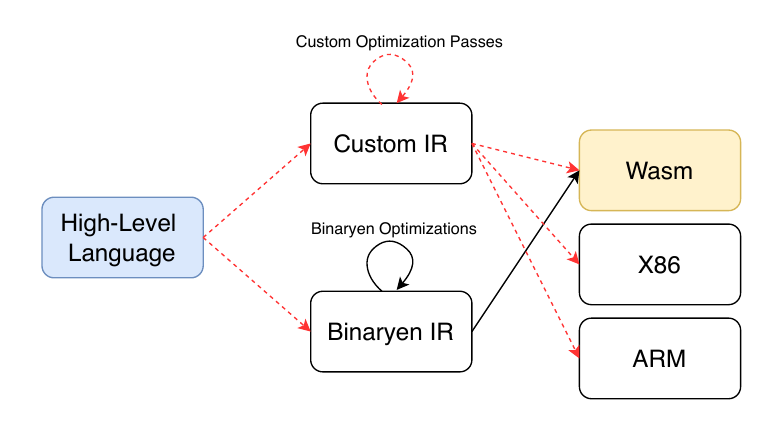}
        \caption{Custom Compiler. Most components (red dotted lines) needs to be
        implemented by the developers of the language compiler.
        }
        \label{fig:custom-compiler}
    \end{subfigure}
    \hfill
    \begin{subfigure}[b]{0.48\textwidth}
        \centering
        \includegraphics[width=\textwidth]{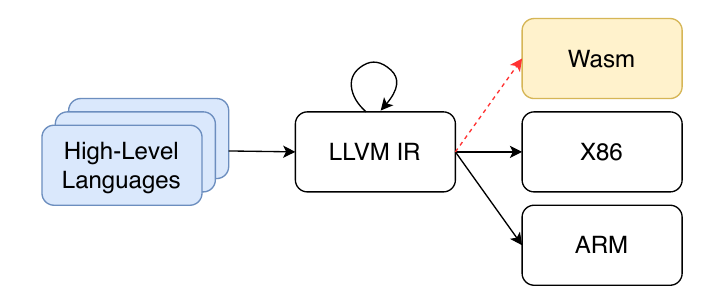}
        \caption{LLVM-based compiler pipeline. 
        LLVM simplifies compiler development by providing components that could
        be reused by different languages and targets.  However, generating Wasm
        from LLVM IR (red dotted line) requires reconstruction of abstractions.
        }
        \label{fig:llvm}
    \end{subfigure}
    
    \vspace{.5em}

    \begin{subfigure}[b]{\textwidth}
    \includegraphics[width=\textwidth]{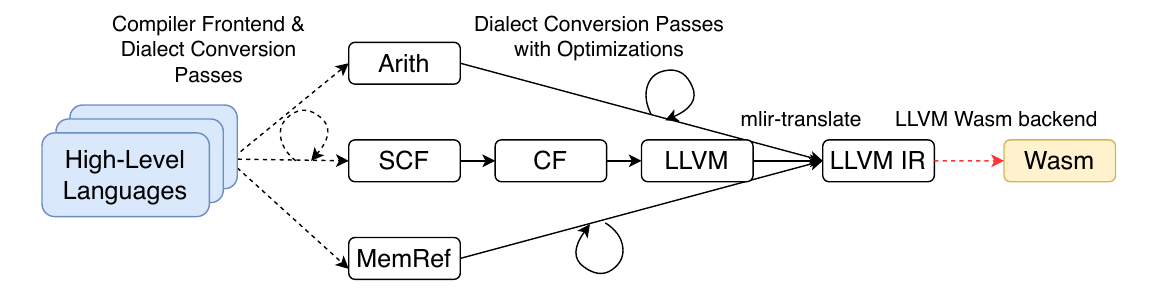}
    \caption{MLIR-based compiler pipeline. 
    MLIR has multiple levels of abstractions
    through dialects,
    but it still relies on LLVM backend for Wasm generation (red dotted line).}
    \label{fig:mlir}
    \end{subfigure}
    \caption{Overview of existing Wasm compilation pipelines.}
    \label{fig:wasm-compiler-landscape}
\end{figure}

\paragraph{Custom Compiler}
Some languages, such as Go, Kotlin, Dart, OCaml, and Java opt to implement their
own compilers for Wasm rather than relying on existing compiler frameworks. 
Figure~\ref{fig:custom-compiler} shows the overview of the custom compiler pipeline.
These compilers can flexibly support Wasm's high-level features, but they
lack reusability and modularity that compiler frameworks like LLVM provide.
Under this approach, developers of compilers for these languages need to implement
compilers by themselves, without taking advantage of
community-contributed optimizations and compilation backends,
which requires a significant engineering effort.
It is challenging for these compilers to maintain up-to-date support for the latest new Wasm features and capabilities.

Some languages, such as AssemblyScript~\cite{assemblyscript} and
Grain~\cite{grain}, use Binaryen~\cite{binaryen} to produce Wasm binaries.
Binaryen is a specialized Wasm compiler and toolchain with its own internal
intermediate representations (IRs), high-level APIs
that language frontends can target,
and a set of optimization passes.
However, because Binaryen is specialized for Wasm, it lacks the modularity and
cross-target reusability found in general compiler frameworks like LLVM.
As a result, languages using Binaryen inherit similar limitations to those built
on custom compiler backends.

\paragraph{LLVM-based Compilers}

Popular languages with Wasm targets, such as C, C++, and Rust, utilize
the LLVM compiler infrastructure~\cite{llvm}.  
Figure~\ref{fig:llvm} shows the overview of the LLVM compiler pipeline.
LLVM uses LLVM IR as an intermediate representation.
Different languages can compile to LLVM IR, and LLVM offers optimizations for
this intermediate representation. 
Subsequently, LLVM IR can be translated to machine code for different targets.  
This modular design enables language implementations to benefit from both LLVM's
optimizations and its diverse backends simply by compiling to LLVM IR.

While the modular design offers advantages, it also introduces an abstraction
loss problem. 
High-level language abstractions vanish during translation to LLVM IR.
This eliminates optimization opportunities that are only available at high-level
abstractions.
Furthermore, this issue is more problematic for Wasm because Wasm often incorporates
high-level instructions that do not have corresponding LLVM IR instructions.
In other words, LLVM IR lacks the expressiveness needed to adequately represent
high-level Wasm features.

To utilize high-level Wasm features in LLVM, one may want to reconstruct the
lost abstractions, but this requires non-trivial engineering efforts, and often infeasible.
For example, since Wasm only allows structured control flow, Emscripten, an
LLVM-to-Wasm compiler, needed to employ a control flow restructuring algorithm 
like Relooper~\cite{emscripten} to convert arbitrary control flow in LLVM IR
into a Wasm-compatible format.
When it comes to Garbage Collection or Stack Switching,
reconstructing the notion of garbage-collected heap objects and non-local
control flow constructs from low-level LLVM IR becomes practically infeasible.

One may want to extend LLVM IR to carry the high-level
information, but this is also nontrivial. 
First of all, LLVM is a large compiler infrastructure, and extending it requires
significant effort and expertise.
While it is relatively easy to extend LLVM to support target-specific behavior
through intrinsics, they are merely functions with predefined semantics, and
therefore not suitable for modeling sophisticated high-level features, such as
non-local control flow.
Attempting to expand LLVM IR's type system is even more daunting, risking
compatibility issues with existing tools \cite{extendingllvm}.
A previous attempt to integrate Wasm GC into LLVM \cite{clangwebasembly}
underscored these challenges, emphasizing the inadequacy of LLVM IR's type
system for capturing Wasm-specific types.

\begin{table}[ht]
    \centering
    \footnotesize
    \begin{tabular}{l c | l c}
    \hline
    \textbf{Proposal} & \textbf{Status} & \textbf{Proposal} & \textbf{Status} \\ 
    \hline
    Tail call & \cellyes & Relaxed SIMD & \cellyes \\
    Extended Const Expressions & \cellyes & Custom Annotation Syntax & \cellna \\
    Garbage Collection & \cellno & Branch Hinting & \cellno \\
    Multiple Memories & \cellno & & \\
    \hline
    \end{tabular}
    \caption{Overview of phase 5 Wasm proposals and their LLVM support.}
    \label{tab:wasm-proposals}
\end{table}

With this fundamental limitation combined with the engineering complexity of
LLVM infrastructure,
LLVM has only limited support for new feature extensions.
Among seven upcoming Wasm features in phase 5, LLVM lacks three (Table
\ref{tab:wasm-proposals}). 
While this limited feature support partly results from low demand among LLVM's
core user base, it also perpetuates a negative feedback loop: compilers needing
high-level Wasm features build separate solutions, which keeps the demand for
such features in LLVM low.
This also impedes existing LLVM-based languages from conditionally using these
features through language extensions~\cite{ref-cpp, cpp-coroutine} or compilation
attributes~\cite{cpp-attributes}, as extending LLVM to support them is both
difficult and often infeasible.

\subsection{MLIR Compiler Infrastructure}

MLIR (Multi-Level Intermediate Representation) is a recently developed compiler
infrastructure maintained by the LLVM community.
Unlike LLVM that relies on a single intermediate representation (LLVM IR),
MLIR introduces a flexible abstraction layer called \emph{dialects}.
These dialects enable a modular and extensible design by providing
tailored intermediate representations for specific languages or domains.
MLIR includes core dialects (Table~\ref{tab:dialects}), including: \tosa for
tensor operations~\cite{tosa}, \memref for memory access~\cite{memref}, \linalg
for linear algebra~\cite{linalg}, \scf for structured control flow~\cite{scf},
\cf for basic control flow~\cite{cf}, and \arith for arithmetic
operations~\cite{arith}.
Each dialect is customized for its respective purpose. 
%
There is also the \llvm dialect, for representing LLVM IR,
which is used to translate MLIR programs to LLVM IR.
In addition, new dialects can be added by developers to support new hardware,
new languages, or new optimizations.

\begin{table}[h]
    \centering
    \begin{tabular}{lll}
    \hline
    \textbf{Dialect} & \textbf{Description} & \textbf{Examples} \\
    \hline
    TOSA  & Tensor operations & \lstinline|tosa.conv2d|, \lstinline|tosa.sigmoid|, \dots \\
    MemRef & Memory-related operations & \lstinline|alloc|, \lstinline|free|, \lstinline|load|, \lstinline|store|, \dots \\
    Arith  & Arithmetic operations & \lstinline|add|, \lstinline|sub|, \lstinline|mul|, \lstinline|divi|, \dots \\
    SCF    & Structured control flow & \lstinline|for|, \lstinline|if|, \lstinline|while|, \dots \\
    CF     & Basic control flow & \lstinline|br|, \lstinline|cond_br|, \dots \\
    LLVM   & Represents LLVM instructions & \lstinline|llvm.add|, \lstinline|llvm.mlir.addressof|, \dots \\
    \hline
    \end{tabular}
    \caption{Some MLIR Dialects and Their Examples}
    \label{tab:dialects}
\end{table}

MLIR has a growing ecosystem.
It is used by numerous machine learning 
and tensor computation compilers
\cite{pienaar2020mlir, zhu2021disc, golin2024towards, jin2020compiling, bik2022compiler},
as well as domain-specific language compilers
\cite{marco, tillet2019triton, rise, mojo}.
By leveraging MLIR, these compilers can streamline the implementation
of new languages and incorporate domain-specific optimizations.
Furthermore, there are active efforts to refactor existing compilers,
either partially or entirely,
to adopt MLIR \cite{fortran-mlir, moses2021polygeist, pylir},
capitalizing on its optimization capabilities, modularity, and extensibility.

Figure~\ref{fig:mlir} illustrates an example pipeline for compiling a high-level
program to Wasm via MLIR, 
a path increasingly adopted by languages through tools such as
Polygeist~\cite{moses2021polygeist} and ClangIR~\cite{clangir}.
A high-level program is first translated into an MLIR representation.
The MLIR program is then incrementally converted into lower-level dialects,
possibly with optimizations at each stage.
Once the program reaches the \llvm dialect, it is converted into LLVM IR 
and subsequently compiled into Wasm via the LLVM backend.
This modular design, composed of multiple dialects and transformation passes,
enables compiler developers to utilize existing dialects and optimizations while
also introducing custom ones tailored to specific compilation requirements.

However, despite its modular architecture and multiple abstraction layers, MLIR
still faces the challenge of abstraction loss inherited from LLVM. 
This issue arises because MLIR relies on LLVM's backend for the final stage of
Wasm compilation. 
Consequently, high-level semantic information is lost during the transition to
LLVM IR, limiting MLIR's ability to effectively support advanced Wasm features
not natively handled by LLVM.

\section{Overview of Our Approach}

\begin{figure}[ht]
    \centering
    \includegraphics[width=\textwidth]{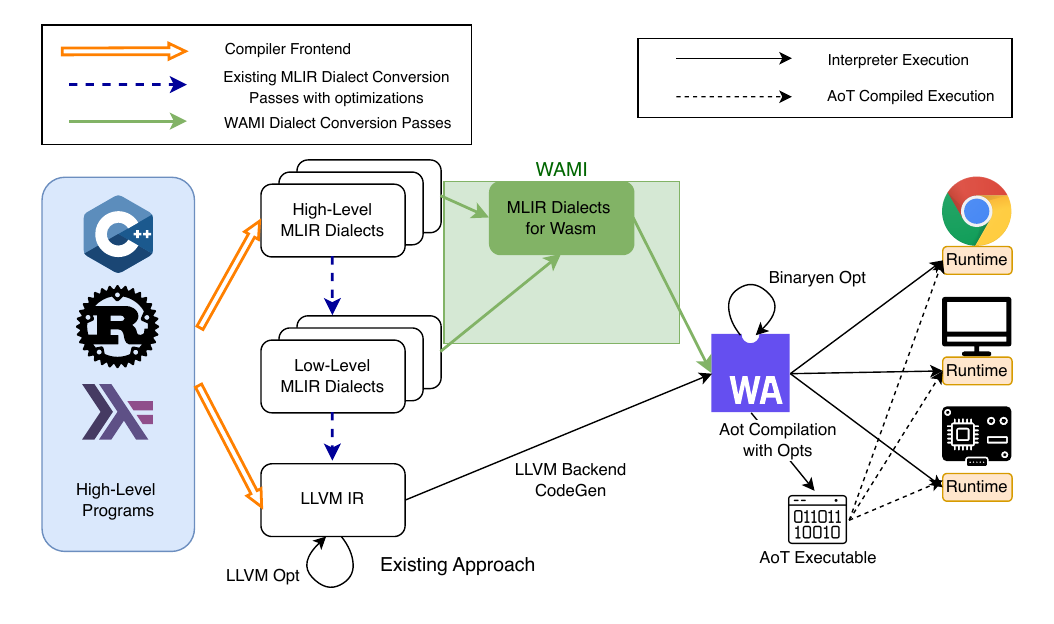}
    \caption{Overview of \tool}
    \label{fig:overview}
\end{figure}

To overcome the limitations associated with abstraction loss in LLVM
and also leverage existing infrastructure for Wasm compilation,
we propose \tool, a novel, MLIR-based compiler pipeline for Wasm.
\tool introduces new MLIR dialects specifically designed to represent Wasm,
which allows high-level MLIR dialects to be converted into Wasm
without loss of abstractions.

Figure~\ref{fig:overview} provides the overview of \tool.
We assume that high-level programs are compiled into MLIR via their
compiler frontends, e.g., Polygeist~\cite{moses2021polygeist} and
ClangIR~\cite{clangir}.
Then, instead of converting the MLIR programs to LLVM IR and using the LLVM
backend to generate Wasm binaries, we directly translate them
into Wasm dialects through dialect conversion passes.
The output MLIR programs are subsequently converted into Wasm
binaries through a simple translation.
The resulting Wasm code 
can execute on any standard
Wasm runtime across various platforms, either through interpretation or via
Just-In-Time (JIT) or Ahead-Of-Time (AoT) compilation to machine code.

\tool have several advantages over existing approaches. 
First, it seamlessly integrates with any language that relies on MLIR as its
compiler infrastructure---a user base expected to grow as MLIR adoption
increases. 
Second, by directly generating Wasm code from high-level MLIR dialects, 
we avoid losing abstractions, allowing high-level Wasm features to be utilized. 
Finally, \tool is designed to be compatible with future Wasm specifications. 
Supporting new Wasm features only requires the modification of the relevant dialect
conversion passes or adding new conversion passes from new dialects that
represent the desired features.

\tool leverages existing optimizations through the pipeline, making it practical
despite bypassing LLVM.
MLIR includes a comprehensive suite of common optimizations such as common
subexpression elimination, dead code elimination, and loop-invariant code
motion. 
Similarly, Binaryen implements a variety of general and Wasm-specific
optimizations, including peephole optimizations and local coalescing.
AoT and JIT compilers in Wasm runtimes also feature their own optimizations,
tailored for the host machine.
These combined optimizations contribute significantly to the pipeline's
performance, enabling results comparable to LLVM-based compilers, as
demonstrated in Section~\ref{sec:evaluation}.

Furthermore, low-level machine code generation algorithms that only exist in
LLVM become less relevant when generating Wasm code.
For example, register allocation is a critical optimization step in most
compilers. 
However, as Wasm does not have a fixed number of physical registers,
sophisticated algorithms involving register
spilling and move insertion are not applicable when generating Wasm code.
The best applicable optimization is to simply minimize the number of virtual
registers, which is already handled by Binaryen.
Additionally, instruction scheduling, another important optimization, offer
little benefit to Wasm code due to the lack of insight into how it will be
executed.
In interpreter-based execution environments,
the concept of instruction-level parallelism becomes irrelevant.
In more sophisticated runtimes, such as TurboFan~\cite{turbofan} and
WAMR~\cite{wamr}, the Wasm code eventually undergoes optimizations including
instruction scheduling through AoT or JIT compilation.
Consequently, although MLIR does not have good support for machine code
generation algorithms~\cite{mlir-documentation},
generating Wasm code from MLIR is practical, without requiring extensive porting
efforts.

\section{Dialect Design}
\label{sec:design}

This section introduces two new MLIR dialects designed to represent Wasm
semantics, creating a direct compilation path from high-level
dialects to Wasm while leveraging MLIR's existing analysis and transformation
capabilities.
The dialects are:
\begin{description}
    \item[\texttt{SsaWasm} dialect] Represents Wasm in Static Single Assignment
        (SSA) form with explicit operands and results,
        facilitating the use of standard MLIR analysis and optimization passes.
    \item[\texttt{Wasm} dialect] Captures Wasm's stack-based semantics
        with implicit operands and results, optimized for direct emission
        of Wasm code. 
\end{description}
The compilation proceeds by lowering high-level dialects into \ssawasm, where
analyses and optimizations can be performed.
A subsequent pass lowers \ssawasm to \wasm by mapping SSA values to the Wasm
execution stack or local variables. 
Finally, the \wasm dialect is translated into Wasm
textual format, which can be converted to Wasm bytecode.
We omit detailed discussion of the \wasm dialect here since it closely mirrors
the semantics and structure of
Wasm.
The remainder of this section details the \ssawasm dialect, focusing on how it
models Wasm instructions, with emphasis on more complex features such as
control flow and Stack Switching related operations.

\begin{figure}[ht]
    \centering
    \begin{tabular}{ll}
        \toprule
        \textbf{Type} & \textbf{Description} \\
        \midrule
        \lstinline|i32|, \lstinline|i64|, \lstinline|f32|, \lstinline|f64| & Builtin numeric types \\
        \lstinline|memref<_>| & Builtin memory reference type, e.g., \lstinline|memref<i32x4>| \\
        \lstinline|!ssawasm.cont<_>| & Continuation reference type, e.g., \lstinline|!ssawasm.cont<"ct">| \\
        \lstinline|!ssawasm.func_ref| & Function reference type \\
        \lstinline|!ssawasm.local<_>| & Local variable type, e.g., \lstinline|!ssawasm.local<i32>| \\
        \bottomrule
    \end{tabular}
    
    \vspace{0.5em}
    
    \setlength{\tabcolsep}{4pt}
    \begin{tabular}{lp{0.8\textwidth}}
        \toprule
        \textbf{Category} & \textbf{Operations} \\
        \midrule
        Arithmetic & \begin{tabular}[t]{@{}l@{}}
                      \lstinline|ssawasm.const|, \lstinline|ssawasm.add|, \lstinline|ssawasm.le_u|, \dots
                    \end{tabular} \\
        \midrule
        Function & \begin{tabular}[t]{@{}l@{}}
                    \lstinline|ssawasm.func|, \lstinline|ssawasm.call|, \lstinline|ssawasm.func_ref|, \dots
                  \end{tabular} \\
        \midrule
        Memory & \begin{tabular}[t]{@{}l@{}}
                  \lstinline|ssawasm.load|, \lstinline|ssawasm.store|, \lstinline|ssawasm.data|, \dots
                \end{tabular} \\
        \midrule
        Local Variables & \begin{tabular}[t]{@{}l@{}}
                           \lstinline|ssawasm.local_decl|, \lstinline|ssawasm.local_set|, \lstinline|ssawasm.local_get|, \dots
                         \end{tabular} \\
        \midrule
        Control Flow & \begin{tabular}[t]{@{}l@{}}
                        \lstinline|ssawasm.block_loop|, \lstinline|ssawasm.block_block|, \lstinline|ssawasm.br|, \dots
                      \end{tabular} \\
        \midrule
        Auxiliary & \begin{tabular}[t]{@{}l@{}}
                     \lstinline|ssawasm.on_stack|, \lstinline|ssawasm.pseudo_br|, \lstinline|ssawasm.pseudo_cond_br|, \\
                     \lstinline|ssawasm.cast_memref_to_i32|, \lstinline|ssawasm.cast_i32_to_memref|
                   \end{tabular} \\
        \bottomrule
    \end{tabular}
    \caption{\ssawasm Types and Operations}
    \label{tab:ssawasm-types-and-operations}
\end{figure}

\subsection{Summary of Types and Operations} 
Table~\ref{tab:ssawasm-types-and-operations} presents the types and operations
used in the \ssawasm dialect. 
The upper table shows the types in used the \ssawasm dialect, which include MLIR
builtin numeric types (\lstinline|i32|, \lstinline|i64|, \lstinline|f32|,
\lstinline|f64|) with the \lstinline|memref<_>| abstraction for structured
memory references.
The dialect also introduces new types (prefixed with
\lstinline|!ssawasm|) that represent Wasm-specific concepts: local variables,
function references, and continuation references (for Stack Switching).
The lower table lists a selection of key operations, categorized by their
semantic domains, from basic arithmetic and memory operations to more complex
control flow constructs.

\paragraph{Modeling Simple Operations}

All Wasm instructions that pop operands from the stack and push the results
back onto the stack are modeled as \ssawasm operations with explicit operands
and results.
For example, the \lstinline|i32.add| operation takes two operands from the
stack, adds them, and pushes the result back onto the stack.
This is modeled as an \lstinline|%result = ssawasm.add %lhs, %rhs : i32|
operation in \ssawasm. 
All arithmetic instructions, memory instructions, such as \lstinline|i32.load|,
and \lstinline|i32.store|, function call instructions, such as \lstinline|call|,
are modeled in this way.

\paragraph{Local Variables}

Local variables in Wasm are modeled as \lstinline|ssawasm.local_decl|
operations, as shown in Fig.~\ref{fig:mlir-local}.
For example, \lstinline|%local = ssawasm.local_decl : !ssawasm.local<i32>| 
declares a local variable \lstinline|%local| of type \lstinline|!ssawasm.local<i32>|.
The new SSA variable, \lstinline|%local|, can be read from and written to via
\lstinline|ssawasm.local_get| and \lstinline|ssawasm.local_set| operations, 
which model the
\lstinline|local.get| and \lstinline|local.set| instructions respectively, in Wasm. 

\begin{figure}[h]
\centering
\begin{subfigure}[t]{0.45\textwidth}
\begin{lstlisting}[style=mlir]
%local= ssawasm.local_decl : !ssawasm.local<i32>
%one = ssawasm.const 1 : i32
ssawasm.local_set %local, %one : i32
%val = ssawasm.local_get %local : i32
\end{lstlisting}
\caption{Usage of \ssawasm local variable-related operations}
\label{fig:mlir-local}
\end{subfigure}
\hfill
\begin{subfigure}[t]{0.5\textwidth}
\begin{lstlisting}[style=mlir]
ssawasm.func @f(%arg1 : ssawasm.local<i32>) -> i32 {
  %val = ssawasm.local_get %arg1 : i32
  return %val : i32
}
ssawasm.func @main() -> () {
  %one = ssawasm.const 1 : i32
  %result = ssawasm.call @f(%one) : (i32) -> i32
}
\end{lstlisting}
\caption{Declaring and calling a function}
\label{fig:mlir-func}
\end{subfigure}

\vspace{.5em}

\begin{subfigure}[t]{\textwidth}
\begin{lstlisting}[style=mlir]
ssawasm.data @data_1 : memref<i32x4> <memory: 0, offset: 1024> = dense<"0XABCDEFGH">
ssawasm.func_import @imported_func : () -> ()
ssawasm.global_var @global_var : i32
\end{lstlisting}
\caption{Module-level operations in \ssawasm}
\label{fig:mlir-module-level}
\end{subfigure}
\end{figure}

\paragraph{Functions}

We use \lstinline|ssawasm.func| to model Wasm functions.
This is analogous to \lstinline|func.func|, but with a different typechecking rule,
as shown by the example in Fig.~\ref{fig:mlir-func}.
All arguments in \lstinline|ssawasm.func| must be of type \lstinline|local<_>|, as
function parameters in Wasm are treated as local variables.
When calling a function, we require the actual arguments to match the types of
the inner types of the parameters,
e.g., \lstinline|i32| instead of \lstinline|local<i32>| in the example.

\paragraph{Module-Level Operations}

We model module-level objects in Wasm, such as data segments, function
imports, and global variable declarations as operations in the \ssawasm dialect.
For example, Fig.~\ref{fig:mlir-module-level} shows the MLIR representation of
a data segment, a function import, and a global variable declaration.
The \lstinline|ssawasm.data| operation contains the memory index and the base
address of the data (1024), and the initial data (\lstinline|dense<"0XABCDEFGH">|).  
It also contains the type of the data (\lstinline|memref<i32x4>|), which is used during
dialect conversion so that other typed memory access operations (e.g.,
\lstinline|memref.load|) can access the type information.

\subsection{Control Flow Operations}

Wasm only supports structured control flow, 
prohibiting arbitrary branching to program points.
Instead, it provides \lstinline|block| and \lstinline|loop| instructions that 
declare labels and have instructions within them.
The branch (\lstinline|br|) and conditional branch (\lstinline|br_if|)
instructions can only branch to one of the labels of the
\lstinline|block| or \lstinline|loop| instructions that enclose them.
Figure~\ref{fig:wasm-loop} illustrates an example of a typical loop in Wasm
using \lstinline|block| and \lstinline|loop| instructions.
The \lstinline|block| and \lstinline|loop| instructions declare labels,
\lstinline|$block_label| and \lstinline|$loop_label|, respectively.
Branching to \lstinline|$loop_label| transfers control
to the beginning of the loop, i.e., line~\ref{line:wasm-loop-begin}.
In contrast, branching to \lstinline|$block_label| transfers control
to the end of the block, i.e., line~\ref{line:wasm-block-end}.

\begin{figure}[bth]
  \centering
  \begin{subfigure}[t]{0.42\textwidth}
      \begin{lstlisting}[style=wasm]
  (block $outer 
    (loop $inner  *@\label{line:wasm-loop-begin}@*
       ;; check termination condition
       ;; branch to end if true
       br_if $outer 
       ...
       ;; update loop vars
       ;; branch to beginning
       br $inner 
    )
  ) *@\label{line:wasm-block-end}@*
  \end{lstlisting}
  \caption{Typical loop in Wasm, represented as a nested \lstinline|block| and \lstinline|loop| instructions}
  \label{fig:wasm-loop}
  \end{subfigure}
  \hfill
  \begin{subfigure}[t]{0.54\textwidth}
      \begin{lstlisting}[style=mlir]
  ssawasm.block_loop {
    ^entry:  // models the code region between block $outer and loop $inner
      ssawasm.pseudo_br ^loop_label
    ^loop_label: // models the loop instruction
      ...
      ssawasm.cond_br %2, ^block, ^b0
    ^b0: // introduced by br_if
      ...
      ssawasm.br ^loop_label *@\label{line:mlir-loop-branch}@*
    ^block_label: // models the block instruction
      ssawasm.exit // no-op
  }
      \end{lstlisting}
      \caption{MLIR representation of a loop using a \lstinline|ssawasm.blockloop| operation}
      \label{fig:ssawasm-loop}
  \end{subfigure}
  \caption{A loop in Wasm and its MLIR representation}
  \label{fig:mlir-loop}
  \end{figure}

Our goal is to model Wasm control flow instructions
in the \ssawasm dialect while preserving their semantics and maintaining
compatibility with MLIR's analysis and transformation passes.
Current LLVM-based Wasm compilers avoid modeling Wasm control flow instructions
directly in MLIR, instead employing control flow restructuring
algorithms~\cite{emscripten, ramsey2022beyond} at a later stage to produce
the \lstinline|block| and \lstinline|loop| instructions of Wasm
from basic blocks and branches in MLIR.
While this approach allows reusing existing compiler infrastructure
for control flow analysis and transformation,
it discards the semantically richer representation--%
such as the \scf dialect--%
by lowering it into basic blocks and branches, only to recover it later.
Although restructuring algorithms perform adequately in most cases, it is an
unnecessarily complex component in compilers for high-level languages that do not have
goto-style control flow.

An alternative approach is to directly generate
Wasm control flow from the high-level dialect (e.g., \scf),
ensuring that the optimal control flow structure is produced from the outset.
This strategy simplifies the compilation pipeline, eases maintainability, and
facilitates seamless integration with other Wasm features involving control
flow, such as stack switching and block arguments.

One challenge is that MLIR, despite its extensibility, 
lacks a direct mechanism for modeling Wasm control flow operations in a
structure-preserving way.
For instance, suppose that we want to model Wasm's \lstinline|block| instruction.
This cannot be modeled as a single basic block in MLIR, as control may be
transferred from the middle of the block to other labels through \lstinline|br|
or \lstinline|br_if| operations, which violates the definition of the basic
block.
If we were to model the \lstinline|block| instruction as multiple basic blocks in MLIR,  
a plausible option would be to model it as an operation with a region, 
which can enclose one or more MLIR basic blocks.
Then, instructions within the \lstinline|block| instruction can be modeled as
operations in one of the basic blocks in the region.
However, this approach does not allow branching to labels
outside their scope, and therefore cannot model Wasm control flow that breaks out of the nested control structures.
For example, the \lstinline|br_if| instruction in Wasm branches to the label of
the \lstinline|block| instruction, which is outside the scope 
of \lstinline|loop| enclosing the \lstinline|br_if|.

To address these challenges, we introduce composite operations that capture the
nested structure of Wasm control flow. 
For example, we designed
\lstinline|ssawasm.block_loop| to represent the nested structure of
\lstinline|block| and \lstinline|loop| instructions, serving as a key building
block for lowering \lstinline|scf.for| and \lstinline|scf.while| to the \ssawasm
dialect (see Section~\ref{sec:dialect-conversion}).
While such composite operations do not surjectively cover all Wasm control flow
patterns, they provide robust building blocks for lowering high-level control
flow in other MLIR dialects, such as \scf.

A \lstinline|ssawasm.block_loop| operation models Wasm code of form
\lstinline[alsoletter={A,B,C},literate={A}{{\textbf{A}\enspace}}1 {B}{{\textbf{B}\enspace}}1 {C}{{\textbf{C}\enspace}}1]|block{ A loop { B } C }|
where the segments \texttt{\textbf{A}}, \texttt{\textbf{B}}, and
\texttt{\textbf{C}} represent (possibly empty) instruction sequences.
Each segment is modeled as one or more basic blocks
within the \lstinline|ssawasm.block_loop| operation's region.
Multiple basic blocks are required to model a single segment
if it contains a \lstinline|br_if| instruction, 
as the code before and after the \lstinline|br_if| need to be modeled as separate
basic blocks.
Branch instructions in Wasm are modeled as terminator operations in MLIR,
which is placed at the end of a basic block and transfers control to other basic
blocks.
We also employ auxiliary operations,
which do not have corresponding Wasm instructions,
but used to explicitly model control flow transfers between basic blocks. 
For example, the implicit control flow transfer from segment \texttt{\textbf{A}} to
segment \texttt{\textbf{B}} in Wasm requires an auxiliary operation in MLIR.
Such operations will be removed by subsequent passes.

Figure~\ref{fig:ssawasm-loop} shows an example of the \lstinline|ssawasm.block_loop|.
The three basic blocks \lstinline|^entry|, \lstinline|^loop_label|, and
\lstinline|^block_label| are predefined by the \lstinline|ssawasm.block_loop| operation,
and the remaining basic block, \lstinline|^b0| 
is introduced while translating the \lstinline|br_if| instruction.
The \lstinline|^entry| block is the entry point of the region, corresponding
to the \texttt{\textbf{A}} segment. 
It contains an auxiliary operation \lstinline|ssawasm.pseudo_br|, which
explicitly transfers control to the \lstinline|^loop_label| block.
The \lstinline|^loop_label| block represents 
the entrypoint of the segment \texttt{\textbf{B}},
and serves as a branch target of the \lstinline|loop| instruction.
Wasm instructions branching to \lstinline|$loop| label
(line~\ref{line:wasm-loop-begin}) are modeled as a terminator operation that
transfer control to the \lstinline|^loop_label| block
(line~\ref{line:mlir-loop-branch}).
The \lstinline|br_if| instruction is modeled as a terminator operation
\lstinline|ssawasm.cond_br|,
that transfers control to either \lstinline|^block_label| or \lstinline|^b0|,
which represents the segment \texttt{\textbf{C}} (the remainder of the loop body).
The second branch target label is an auxilary value used to mark the control flow,
and will be removed by the dialect conversion pass.
The \lstinline|^block_label| block 
represents the branch target of the \lstinline|block| instruction.
It contains no operations and is used to end the region.

\subsection{Auxiliary Operations}
In addition to the \lstinline|ssawasm.pseudo_br| operation, we introduce
additional auxiliary operations that do not have corresponding Wasm instructions
to facilitate dialect conversion passes and are removed in the final lowering
pass to Wasm.

The \lstinline|ssawasm.on_stack| operation is used to model values that are implicitly pushed onto the stack.  
For example, the \lstinline|block| instructions in Wasm
may have return values.  
When a value is returned from a \lstinline|block|, it is implicitly pushed onto the
stack when control transfers to the end of the \lstinline|block|.  
We model such a value as follows:
\lstinline|%value = ssawasm.on_stack : i32|. 
Other operations can explicitly reference the value \lstinline|%value|.

We also introduce an auxiliary type casting operation,
to aid dialect conversion from \memref. 
The \memref dialect provides an abstraction for type-safe memory
operations.
For instance, \memref values have types with a specific shape, e.g.,
\lstinline|memref<i32x10x10>|, and elements can be accessed through
indices, e.g., \lstinline|%elem = memref.load %memref[%idx0, %idx1]|.
To support seamless conversion from the \memref operations to \ssawasm
operations, we preserve the rich type information of \memref
throughout the dialect conversion,
rather than immediately flattening to raw address types (\lstinline|i32| or
\lstinline|i64|).
This allows downstream compiler passes to leverage the shape information
to compute memory offset and size.
We extend the \ssawasm dialect to support memref-typed
results in constants and function calls representing memory addresses.
Additionally, we introduce an auxiliary type casting operation,
\lstinline|ssawasm.cast_memref_to_i32|, which converts \memref types to address
types, when needed by subsequent \ssawasm operations.
These casting operations provide a clean separation of concerns and are
systematically removed during the final dialect conversion passes.

\setlength{\columnsep}{3em}
\begin{figure}[tbp]
\begin{multicols}{2}
\begin{lstlisting}[style=wasm, firstnumber=1]
type $ft (func (result i32)) *@\label{line:wasm-ft}@*       
type $ct ($ft) *@\label{line:wasm-ct}@*                  
tag $yield (param i32) *@\label{line:wasm-yield}@*
(func $task (result i32) 
  i32.const 1
  suspend $yield ;; suspend the continuation with value 1
  i32.const 2
  return
)
\end{lstlisting}
\setcounter{lstnumber}{9}
\columnbreak
\begin{lstlisting}[style=wasm, firstnumber=10]
(func $main 
  (block $on_yield (result (ref $ct) i32) 
    ref.func $task         
    cont.new $ct           
    resume $ct (on $yield $on_yield) 
    ;; stack: [2]
  )
  ;; stack: [ref to continuation, 1]
)
\end{lstlisting}
\end{multicols}
\vspace{-10pt}
\caption{Example of Stack Switching instructions in Wasm}
\label{fig:wasm-stack-switching}
\end{figure}
\subsection{Stack Switching}

To demonstrate the flexibility of the \ssawasm dialect, we model the early-stage
feature, Stack Switching.
Figure~\ref{fig:wasm-stack-switching} shows an example usage of Stack Switching.
On lines \ref{line:wasm-ft}--\ref{line:wasm-yield}, we define a continuation
type derived from a function type and a tag.
Tags generalize exceptions, allowing a continuation to raise them when
suspended. 
The parent, which resumed the continuation, can catch them.
Tags may have arguments and results, which are types of values that can be passed
to and from between the continuation and the parent.
In the \lstinline|main| function, 
the \lstinline|resume| operation consumes a continuation---%
created via \lstinline|cont.new| operation
from a \lstinline|ref.func $task| reference---%
and transfers control to it.
The handler installed in the \lstinline|resume| operation
(\lstinline|on $yield $on_yield|)
redirects control to the suspend handler,
which is \lstinline|$on_yield|,
if the continuation suspends with the \lstinline|$yield|
tag (e.g., \lstinline|suspend $yield| in the \lstinline|task| function).
If the continuation completes without being suspended,
the \lstinline|resume| operation transfers control to the next operation
in the \lstinline|main| function,
returning the value returned by the continuation.

Most Stack Switching instructions map cleanly to the \ssawasm dialect.
The \lstinline|cont.new| instruction corresponds to \lstinline|ssawasm.cont_new|,
which consumes a function reference---created via \lstinline|ssawasm.func_ref|---%
and produces a fresh continuation reference.
Similarly, the \lstinline|suspend| instruction maps to \lstinline|ssawasm.suspend|,
which accepts values matching the tag's input argument types and returns a value of the type specified by the
tag's result arguments.

The \lstinline|resume| instruction, which involves control flow transfer
through suspend handlers, is modeled as shown in Figure~\ref{fig:ssawasm-resume}.
Currently, we only support modeling the \lstinline|resume| instruction
with a single handler.
It takes arguments of types (input arguments of the continuation type) and a
reference to the continuation that is being resumed,
and returns values (result arguments of the continuation type).
The suspend handler is represented as a block reference \lstinline|^on_yield|,
which could point to a basic block in a \ssawasm control flow construct,
such as \lstinline|ssawasm.block_loop|.
Since the \lstinline|ssawasm.resume| operation transfers control,
it must be the last operation in its basic block,
and therefore subsequent operations are placed in a new basic block,
linked to the \lstinline|resume| operation via a reference (\lstinline|^resume|).

\begin{figure}[h]
    \begin{lstlisting}[style=mlir]
    %return = ssawasm.resume(%arg1, ..., %argN, %cont)[^on_yield, ^resume] @yield %cont : (!ssawasm.cont<ct>) -> (i32)
^resume: // new block created. following operations are located in this block
    \end{lstlisting}
    \caption{The \lstinline|resume| operation in \ssawasm}
    \label{fig:ssawasm-resume}
\end{figure}

\section{Dialect Conversion Passes}
\label{sec:dialect-conversion}

In this section, we describe our conversion passes from core MLIR dialects to \ssawasm. After obtaining the \ssawasm dialect, we convert it to the
\wasm dialect, which directly mirrors Wasm's textual format, simplifying the
generation of Wasm text.

\subsection{Core Dialects to SsaWasm}
We currently support dialect conversion from \arith, \memref, \func, and \scf
dialects to the \ssawasm dialect, and this list remains flexible and
extensible.
We choose dialects for conversion based on their abstraction level and targeted
Wasm features. For example, we convert \scf directly to \ssawasm, instead of
first compiling \scf to \cf and then implementing a \cf-to-\ssawasm pass, to
avoid the need for a control flow restructuring algorithm.

The new dialect conversion passes we introduce are:
\lstinline|arith-to-ssawasm|,
\lstinline|memref-to-ssawasm|,
\lstinline|func-to-ssawasm|,
\lstinline|scf-to-ssawasm|,
and \lstinline|ssawasm-to-wasm|.
High-level dialects can first be converted to \arith, \memref, \func, and \scf
dialects by existing conversion passes, and then converted to \ssawasm dialect
through these four passes.
Then, the \ssawasm dialect can be converted to the Wasm dialect through
\lstinline|ssawasm-to-wasm|.

\paragraph{\lstinline|arith-to-ssawasm| and \lstinline|func-to-ssawasm|}
These two passes transform \arith and \func operations into their \ssawasm
equivalents.
The operands and results remain the same,
and only the opcode is replaced with the corresponding \ssawasm operation.
For example, during the conversion,
\lstinline|%result = arith.addi %lhs, %rhs : i32| is 
replaced with the following \ssawasm instruction
\lstinline|%result = ssawasm.add %lhs, %rhs : i32|.

\paragraph{\lstinline|memref-to-ssawasm|}

\begin{figure}[h]
\centering
\begin{minipage}[t]{0.47\textwidth}
\begin{subfigure}[t]{\textwidth}
    \begin{lstlisting}[style=mlir]
func.func @main() {
  %memref = memref.alloc() : memref<i32x4>
  %index = arith.constant 1 : i32
  %result = memref.load %memref[%index] : memref<i32x4>
}
\end{lstlisting}
\caption{Example usage of \lstinline|memref.alloc| and \lstinline|memref.load|}
\label{fig:memref-alloc-to-ssawasm-before}
\end{subfigure}

\vspace{1em}

\begin{subfigure}[t]{\textwidth}
\begin{lstlisting}[style=mlir]
memref.global @data_1 : memref<i32x4> = dense<"0XABCDEFGH">
memref.global @data_2 : memref<i32x4> = dense<"0X12345678">
func.func @main() {
 %memref = memref.get_global @data_1 : memref<i32x4>
}
\end{lstlisting}
\caption{Example usage of \lstinline|memref.global| and \lstinline|memref.get_global|}
\label{fig:memref-global-to-ssawasm-before}
\end{subfigure}
\end{minipage}
\hfill
\begin{minipage}[t]{0.47\textwidth}
\begin{subfigure}[t]{\textwidth}
\begin{lstlisting}[style=mlir]
func.func @main() {
  // size to alloc computed based on the memref type
  %size = ssawasm.const 16 : i32
  %base_addr = ssawasm.call @malloc(%size) : (i32) -> memref<i32x4> *@\label{line:memref-alloc-to-ssawasm-malloc}@*
  %index = ssawasm.const 1 : i32
  %stride = ssawasm.const 4 : i32
  %mul = arith.muli %index, %stride : i32
  %base_addr_i32 = ssawasm.cast_memref_to_i32 %base_addr : memref<i32x4> -> i32 *@\label{line:memref-alloc-to-ssawasm-cast-memref-to-i32}@*
  %addr = ssawasm.add %base_addr_i32, %mul : i32
  %result = ssawasm.load %addr : i32
}
\end{lstlisting}
\caption{Conversion of Fig.~\ref{fig:memref-alloc-to-ssawasm-before} to \ssawasm}
\label{fig:memref-alloc-to-ssawasm-after}
\end{subfigure}
\end{minipage}

\vspace{.5em}

\begin{subfigure}[t]{\textwidth}
\begin{lstlisting}[style=mlir]
// 1024 is the default start address for data segments in Wasm
ssawasm.data @data_1 : memref<i32x4> <memory: 0, offset: 1024> = "....";
ssawasm.data @data_2 : memref<i32x4> <memory: 0, offset: 1040> = "....";
func.func @main() {
  %memref = ssawasm.const 1024 : memref<i32x4>
  %addr = ssawasm.cast_memref_to_i32 %memref : memref<i32x4> -> i32
  // all uses of %memref are replaced with %addr
}
\end{lstlisting}
\caption{Conversion of Fig.~\ref{fig:memref-global-to-ssawasm-before} to \ssawasm}
\label{fig:memref-global-to-ssawasm-after}
\end{subfigure}
\caption{Examples of conversion from \memref to \ssawasm}
\label{fig:memref-to-ssawasm}
\end{figure}

We support memory operations by providing dialect conversion from \memref to \ssawasm.
The \lstinline|memref.alloc| and \lstinline|memref.dealloc|, which allocates and
deallocates heap memory, are converted to
\lstinline|ssawasm.call @malloc| and \lstinline|ssawasm.call @free|, the same as the LLVM backend. 
We link these library functions after generating the Wasm binary.
As shown on line~\ref{line:memref-alloc-to-ssawasm-malloc} in
Figure~\ref{fig:memref-alloc-to-ssawasm-after}, \lstinline|ssawasm.call @malloc|
returns a value of type \lstinline|memref<i32x4>|. 
It is converted to \lstinline|i32|
(line~\ref{line:memref-alloc-to-ssawasm-cast-memref-to-i32})
and used as the base address of the allocated memory.

We leave \lstinline|memref.alloca|, which allocates memory on the stack, as a
future work.
This can be implemented by modeling the stack in the conversion pass and adding
operations for allocation and deallocation based on scope.

\lstinline|memref.load| and \lstinline|memref.store| are lowered to
\lstinline|ssawasm.load| and \lstinline|ssawasm.store|, respectively, with
address computations explicitly decomposed into primitive arithmetic operations
(multiplying strides by indices, adding offsets to the base address, etc), as
shown in Figure~\ref{fig:memref-alloc-to-ssawasm-after}. 
Unlike LLVM IR, which uses the
\lstinline|GetElementPtr| instruction to abstract address computation from
memory access operations, our approach explicitly incorporates address
arithmetic operations before load/store operations.
In the MLIR ecosystem, the higher-level \memref dialect already serves the abstraction
role that \lstinline|GetElementPtr| plays in LLVM IR, providing type-safe structured 
memory access without obscuring the underlying address computation.
By decomposing address computations into explicit arithmetic in \ssawasm, we
expose these operations to MLIR's optimization passes, enabling common
subexpression elimination, constant folding, and other transformations to be
applied to address computations. 
This strategic decision helps compensate for the absence of specialized machine
code generation optimizations in MLIR.

The \lstinline|memref.global| operation, which declares global data, is lowered to
\lstinline|ssawasm.data| (as shown in Fig.~\ref{fig:memref-global-to-ssawasm-before} and Fig.~\ref{fig:memref-global-to-ssawasm-after}).
Each \lstinline|ssawasm.data| is assigned a memory index and a base address,
which is computed according to the data's size, alignment, and offset.
The \lstinline|memref.get_global| operation, which references a global value, is
converted into \lstinline|ssawasm.const| with the value being the base address
of the \lstinline|ssawasm.data|.

\paragraph{\lstinline|scf-to-ssawasm|}

\begin{figure}[h]
\begin{subfigure}[t]{\textwidth}
\begin{lstlisting}[style=mlir]
%sum = scf.for %i = %lb to %ub step %step iter_args(%acc = %zero_i32) -> (i32) {
    %new_acc = arith.addi %acc, %num : i32
    scf.yield %new_acc : i32
}
\end{lstlisting}
\caption{\lstinline|scf.for| loop before conversion}
\label{fig:scf-to-ssawasm-before}
\end{subfigure}

\vspace{.5em}

\begin{subfigure}[t]{\textwidth}
\begin{lstlisting}[style=mlir]
%acc = ssawasm.local_decl : !ssawasm.local<i32>
%i = ssawasm.local_decl : !ssawasm.local<i32>
%result = ssawasm.local_decl : !ssawasm.local<i32>
ssawasm.block_loop {
    ^entry // initialize the loop variables
        ssawasm.local_set %acc, %zero_i32 : i32
        ssawasm.local_set %i, %lb : i32
    ^loop_label // serves as the loop instruction in Wasm
        %cond = arith.cmpi eq, %i, %ub : i32
        ssawasm.pseudo_cond_br %cond, ^body, ^block_label : i32
    ^body // body of scf.for is translated here
        %acc_val = ssawasm.local_get %acc : i32
        %num_val = ssawasm.local_get %num : i32
        %new_acc = arith.addi %acc_val, %num_val : i32
        ssawasm.pseudo_br ^ind_var_update
    ^ind_var_update // update the loop induction variable
        %new_i = arith.addi %i, %step : i32
        ssawasm.local_set %i, %new_i : i32
        ssawasm.local_set %result, %new_acc : i32
        ssawasm.pseudo_br ^loop_label
    ^block_label // serves as the block instruction in Wasm
        ssawasm.exit // no-op
}
\end{lstlisting}
\caption{\lstinline|scf.for| loop converted to \lstinline|ssawasm.block_loop|}
\label{fig:scf-to-ssawasm-after}
\end{subfigure}
\caption{Example of conversion from \scf to \ssawasm}
\label{fig:scf-to-ssawasm}
\end{figure}

We transform control flow operations from the \scf dialect to the \ssawasm
dialect using the composite operations described in Section~\ref{sec:design}.
Figure~\ref{fig:scf-to-ssawasm} illustrates this transformation with a simple
\lstinline|scf.for| loop example. 
Both the induction
variable (\lstinline|%i|) and loop-carried variable (\lstinline|%acc|) are
converted to local variables.
The \lstinline|^loop_label| block handles condition checking. 
Control transfers either to the \lstinline|^body| block when the
condition evaluates to true or to the \lstinline|^block_label| block when the condition evaluates to false.
Within the \lstinline|^body| block,
which contains the original \lstinline|scf.for| loop contents, all variable
accesses are replaced with references to the previously declared local
variables. 
At completion, the \lstinline|^body| block transfers control to the
\lstinline|^ind_var_update| block, which updates both the induction and
loop-carried variables before redirecting control back to the
\lstinline|^loop_label| block to reevaluate the condition.

\subsection{\ssawasm to \wasm}
The \ssawasm to \wasm transformation implements an MLIR pipeline, which is
essentially a simplified version of the Wasm code generation implemented in the
LLVM backend.
It is divided into three sub-passes: \lstinline|ssawasm-global-to-wasm|,
\lstinline|introduce-locals|, and \lstinline|ssawasm-to-wasm|.

The \lstinline|ssawasm-global-to-wasm| pass transforms global
operations (i.e., those defined outside of functions) into their Wasm
counterparts. For example, \lstinline|ssawasm.data| operations become
\lstinline|wasm.data|, with their content encoded as UTF-8 text for the textual
Wasm format. 

\begin{figure}[htbp]
\centering
\begin{minipage}[t]{0.5\textwidth}
    \centering
    \begin{subfigure}[t]{\textwidth}
    \begin{lstlisting}[style=mlir]
%0 = ssawasm.const 1 : i32
%1 = ssawasm.const 2 : i32
%2 = ssawasm.add %0, %1 : i32
%3 = ssawasm.add %0, %2 : i32
    \end{lstlisting}
    \caption{\ssawasm code without local variables.}
    \label{fig:introduce-locals-before}
    \end{subfigure}\\[1em]
    
    \renewcommand\thesubfigure{c}
    \begin{subfigure}[t]{\textwidth}
    \begin{lstlisting}[style=mlir]
wasm.block "block_label" {
wasm.loop "loop_label" {
    ...
    wasm.br_if "block_label"
    ...
    wasm.br "loop_label"
    }
}
    \end{lstlisting}
    \caption{A \lstinline|ssawasm.block_loop| in Fig.~\ref{fig:scf-to-ssawasm-after} converted to the \wasm dialect.}
    \label{fig:convert-ssawasm-to-wasm-after}
    \end{subfigure}
\end{minipage}%
\hfill
\begin{minipage}[t]{0.45\textwidth}
    \centering
    \renewcommand\thesubfigure{b}
    \begin{subfigure}[t]{\textwidth}
    \begin{lstlisting}[style=mlir]
%0_local = ssawasm.local_decl
%1 = ssawasm.const 1 : i32
ssawasm.local_set %0_local, %1
%0_1 = ssawasm.local_get %0_local
%2 = ssawasm.const 2 : i32
%3 = ssawasm.add %0_1, %2
%0_2 = ssawasm.local_get %0_local
%4 = ssawasm.add %0_2, %3
    \end{lstlisting}
    \caption{The result of applying \lstinline|introduce-locals| to the \ssawasm code snippet in (a).}
    \label{fig:introduce-locals-after}
    \end{subfigure}
\end{minipage}

\caption{Examples of conversion from \ssawasm to \wasm}
\label{fig:ssawasm-to-wasm}
\end{figure}

The \lstinline|introduce-locals| pass introduces local variables
when values cannot be passed directly through the execution stack (i.e., pushed
and popped).
As an example, Figure~\ref{fig:introduce-locals-before} shows value
\lstinline|%0| being referenced multiple times, making stack-only passing
impossible.
By introducing a local variable \lstinline|%0_local|, the value of
\lstinline|%0| can be stored, and subsequent uses of \lstinline|%0| are replaced
with values read from \lstinline|%0_local| using \lstinline|ssawasm.local_get|. 
This design parallels the \lstinline|explicit-locals| pass in the LLVM Wasm
backend.

The \lstinline|ssawasm-to-wasm| pass performs the dialect
conversion from \ssawasm to the \wasm dialect by mapping each \ssawasm operation
to a corresponding \wasm operation. Any operands and results on \ssawasm
operations are dropped, assuming \lstinline|introduce-locals| has inserted the
necessary local variables related instructions. Auxiliary casting operations are
discarded as well.

Control flow is converted by mapping \lstinline|ssawasm.block_loop| to a nested
\lstinline|wasm.block| and \lstinline|wasm.loop| structure. These operations
include labels and single-block regions; branching is represented via
\lstinline|wasm.br| or \lstinline|wasm.br_if| instructions. Unlike standard MLIR
control flow operations, \lstinline|wasm.br| and \lstinline|wasm.br_if| are not
terminator operations. They instead serve as simple markers with target labels,
so that they can be replaced with the corresponding Wasm operation.
Figure~\ref{fig:convert-ssawasm-to-wasm-after} illustrates how
\lstinline|ssawasm.block_loop| is lowered into nested Wasm constructs.
Auxiliary operations (e.g., \lstinline|ssawasm.pseudo_br|) are removed during the conversion.

\section{Case Study: Compilation to Wasm with Stack Switching}
\label{sec:casestudy}

In this section, we show an example of compiling a high-level MLIR dialect to
Wasm through \tool,
leveraging high-level features by preserving the abstraction in the dialect.
To demonstrate the approach, we introduce \dcont, a simple dialect 
that models delimited continuations,
which is a programming model for representing non-local control flow.
We provide examples that illustrate the conversion from the \dcont dialect to the
\ssawasm dialect, leveraging the Stack Switching feature,
showing how \tool can be used to compile high-level MLIR dialects to Wasm.

\subsection{Delimited Continuation Dialect}

As support for dialects with non-local control flow in MLIR is not yet mature, 
we begin by introducing a prototype dialect called \dcont (delimited
continuation), along with an associated dialect conversion pipeline for
demonstration purposes.
\dcont is designed to be a compilation target for higher-level languages and
dialects to model non-local control flow.
Compared to the \async dialect, \dcont dialect is more flexible and supports more
general non-local control flow patterns such as generators.  The \async dialect
can also be lowered to \ssawasm, though it is out of the scope of this work.

Delimited continuations~\cite{felleisen1988theory} provide a programming model
featuring first-class control operators capable of capturing and manipulating
execution contexts.
This capability enables sophisticated control flow mechanisms such as green
threads and yield-style generators by allowing programmers to suspend, resume,
and compose execution contexts dynamically.
Notably, Wasm's stack-switching proposal adopts this delimited
continuation model.

\begin{figure}[ht]
\begin{lstlisting}[style=mlir]
func.func @generator() {
    ... // c0, c1, c1000 are constants of type index
    scf.for %i = %c0_index to %c1000_index step %c1_index {
      %val = arith.index_cast %i : index to i32
      dcont.suspend (%val) : (i32) -> () *@\label{line:dcont-suspend}@*
    }
  return
}
func.func @main() {
  %task1_handle = dcont.new @generator : !dcont.cont<(i32)->()> *@\label{line:dcont-new}@*
  %storage = dcont.alloc() : !dcont.allocated<(i32)->()>
  dcont.store %storage, %task1_handle : !dcont.cont<(i32)->()> 
  ... // c0, c1, c1000 are constants of type index
  scf.for %i = %c0_index to %c1000_index step %c1_index {
    %loaded = dcont.load %storage : !dcont.cont<(i32)->()>
    dcont.resume(%loaded) *@\label{line:dcont-resume}@*
      {(%val, %suspended_cont: i32, !dcont.cont<(i32)->()>): 
        // print %val
        dcont.store %storage, %suspended_cont : !dcont.cont<(i32)->()> 
      }
  }
  return
}

\end{lstlisting}
\caption{A simple generator example in the \dcont dialect}
\label{fig:dcont-simple-example}
\end{figure}

Figure~\ref{fig:dcont-simple-example} illustrates the operations in
the \dcont dialect through a simple example of yield-style generator. 
The \lstinline|dcont.new| operation (line~\ref{line:dcont-new}) creates a new
continuation from a specified function (\lstinline|generator|) of type
\lstinline|() -> ()|. 
The generated continuation has type \lstinline|!dcont.cont<(i32)->()>|, 
indicating it yields a value of type \lstinline|i32| to the parent.
Memory allocation for this continuation is managed by the
\lstinline|dcont.alloc| operation. The continuation is stored to using
\lstinline|dcont.store|.
Within the \lstinline|main| function's loop, the continuation is retrieved from
memory with \lstinline|dcont.load| and resumed via \lstinline|dcont.resume|.
Upon resumption, the continuation executes until it encounters a
\lstinline|dcont.suspend| operation (line~\ref{line:dcont-suspend}).
When suspension occurs, control transitions to the region within \lstinline|dcont.resume|,
which is the suspend handler installed by the operation.
This handler gains access to both the suspended continuation and the values
passed from the \lstinline|dcont.suspend| operation (e.g., \lstinline|%val| on
line~\ref{line:dcont-suspend}).
The handler prints the yielded value and stores the suspended continuation back
into memory allocated earlier by \lstinline|dcont.alloc|.

\subsection{Dialect Conversion}

We support dialect conversion from the \dcont dialect to the \ssawasm dialect.
The \lstinline|dcont.new| operation is converted to a sequence of
\lstinline|ssawasm.func_ref| and \lstinline|ssawasm.cont_new| operations.
The \lstinline|dcont.suspend| operation
is converted to \lstinline|ssawasm.suspend|
operation.
The \lstinline|dcont.alloc|,
\lstinline|dcont.load|, and \lstinline|dcont.store| operations
are converted to local variable declaration,
load, and store operations, respectively.

\begin{figure}[ht]
\begin{subfigure}[c]{\textwidth}  
\begin{lstlisting}[style=mlir]
dcont.resume %cont : (dcont.cont<(i32)->()>, i32) -> ()
{(%val, %suspended_cont: i32, !dcont.cont<(i32)->()>) -> () {
    ... // handler code ...
}}
\end{lstlisting}
\caption{Before conversion: \lstinline|dcont.resume|}
\end{subfigure}

\vspace{.5em}

\begin{subfigure}[c]{\textwidth}  
\begin{lstlisting}[style=mlir]
ssawasm.block_block {
^entry:
    ssawasm.resume %cont (on_yield:^inner_block_label,fallback:^resume)
^resume:
    ssawasm.pseudo_br ^outer_block_label
^inner_block_label:
    %suspended_cont = ssawasm.on_stack : ssawasm.cont<(i32)->()>
    %val =  ssawasm.on_stack : i32
    ... // handler code ...
    ssawasm.pseudo_br ^outer_block_label
^outer_block_label:
}
\end{lstlisting}
\caption{After conversion: \lstinline|ssawasm.resume| within \lstinline|ssawasm.block_block|}
\end{subfigure}
\caption{Conversion of \lstinline|dcont.resume|}
\label{fig:dcont-resume}
\end{figure}

To support dialect conversion of \lstinline|dcont.resume|, we introduce
\lstinline|ssawasm.block_block| to represent a nested structure 
\lstinline[alsoletter={A,B,C},literate={A}{{\textbf{A}\enspace}}1 {B}{{\textbf{B}\enspace}}1 {C}{{\textbf{C}\enspace}}1]|block{ A block { B } C }|
similar to the \lstinline|ssawasm.block_loop| operation.
Figure~\ref{fig:dcont-resume} shows an example of the conversion of
\lstinline|dcont.resume|.
The \lstinline|dcont.resume| operation is converted to 
a \lstinline|ssawasm.resume| operation
within a \lstinline|ssawasm.block_block| operation,
which is used to model the control flow involving suspend handlers.
The \lstinline|^entry| block represents the \texttt{\textbf{A}} segment,
and include a call to \lstinline|ssawasm.resume| operation.
If the continuation executed by \lstinline|ssawasm.resume| yields,
the control transfers to the \lstinline|^inner_block_label| block,
as set by the \lstinline|on_yield| argument of \lstinline|ssawasm.resume|.
The \lstinline|^inner_block_label| block
contains the \lstinline|on_yield| handler of \lstinline|ssawasm.resume|,
where the implicit arguments of the handler
are represented as temporary operations \lstinline|ssawasm.on_stack|.
If the continuation executed by \lstinline|ssawasm.resume| 
finishes execution without yielding,
the control transfers to the \lstinline|^resume| block,
as set by the \lstinline|fallback| argument of \lstinline|ssawasm.resume|.
The \lstinline|^resume| block transfers control to the
\lstinline|^outer_block_label| block, exiting the
\lstinline|ssawasm.block_block| operation.

We use two simple scenarios as examples: a typical yield-style generator
in Fig.~\ref{fig:dcont-simple-example} and a cooperative task scheduler.
The cooperative task scheduler example implements two tasks concurrently process
different halves of a shared global array, doubling each value, with task
switching occurring after each array element modification.
The MLIR programs may have operations from multiple dialects along with \dcont,
e.g., \scf, \func, and \arith.
The dialect conversion pipeline of \tool can convert them to \ssawasm
in a modular fashion, by applying the conversion passes
\lstinline|dcont-to-ssawasm|, \lstinline|scf-to-ssawasm|, \lstinline|arith-to-ssawasm|.  
We confirmed that the generated Wasm code runs correctly using
Wasmfxtime~\cite{wasmfxtime}, a reference implementation of Stack Switching. 

\subsection{Discussion}

The examples in this section demonstrate \tool's ability to seamlessly
convert dialects with non-local control flow to Wasm with the Stack Switching
feature.
While LLVM coroutine intrinsics~\cite{llvm-coroutines} can achieve similar
functionality, Stack Switching offers distinct advantages.
For instance, future integration with the JavaScript MLIR dialect~\cite{jsir}
could provide more efficient implementations of Wasm functions interacting with
JavaScript promises, eliminating the overhead currently imposed by Asyncify~\cite{asyncify}.
Furthermore, Stack Switching provides a better abstraction compared to
LLVM-based approaches that implement stack switching within Wasm. 
For example, on microcontrollers, the runtime that executes Wasm code can
optimize how the Stack Switching instructions are handled based on the hardware
features of the microcontroller---%
such as by utilizing specific memory regions for saving and restoring the stack,
or selectively disabling interrupts during the switching, without losing the
portability of the Wasm code.
\section{Performance Evaluation}
\label{sec:evaluation}

In this section, we evaluate the performance of code generated by our 
Wasm dialect-based compiler against LLVM-generated code using the PolyBench
benchmark suite on a MacBook M2 Pro and an Apollo4 Blue Plus microcontroller,
using both interpreters and AoT compilers. 

\subsection{Implementation and Test Setup}

We implement \tool as an out-of-tree dialect and conversion pass collection for
MLIR.
The implementation, including the \dcont dialect and its conversion
passes, comprises approximately 4,800 lines of C++ and TableGen~\cite{tablegen}
code. 
This breaks down into 1,900 lines for the \ssawasm and \wasm dialects, and
2,900 lines for dialect conversion passes—including the final conversion from
the \wasm dialect to Wasm textual format. 
The entire system was developed by a single developer over a five-month period.

Our evaluation uses MLIR code derived from the PolyBench C benchmark
suite~\cite{polybench} via Polygeist~\cite{moses2021polygeist}.
PolyBench encompasses numerical computations commonly seen in on-device machine
learning workloads, such as linear algebra and image processing kernels.
We evaluated 29 out of 30 available benchmarks; the \lstinline|deriche|
benchmark was excluded due to compilation issues with Polygeist.
Additionally, tests including \lstinline|durbin|, \lstinline|gramschmidt|,
\lstinline|ludcmp|, and \lstinline|symm|, involve stack allocation via
\lstinline|memref.alloca|, for which \tool lacks full support.
For these, we treat them as heap allocation (via \lstinline|memref.alloc|) in
\tool.
This does not threaten the validity of our results, as the performance of the
heap allocation is slower than the stack allocation.

\begin{figure}[h]
\centering
\includegraphics[width=\textwidth]{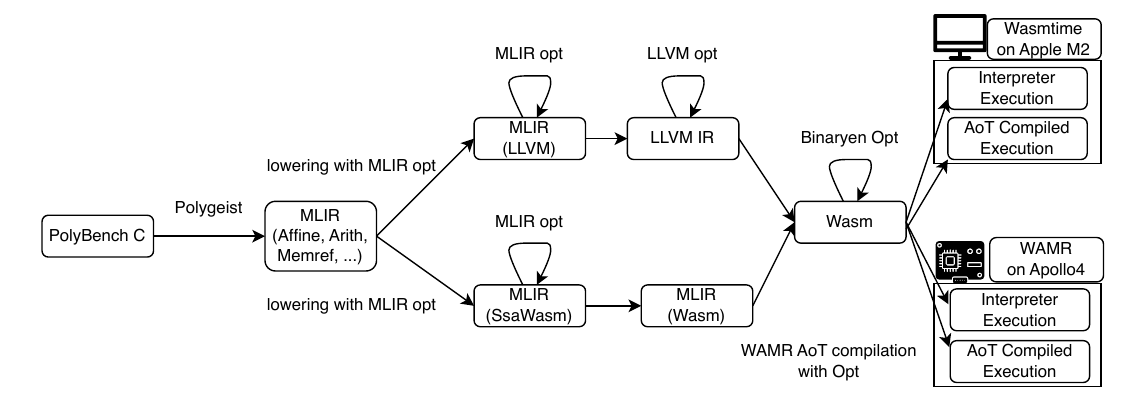}
\caption{Compilation and execution pipelines for evaluation.}
\label{fig:compilation-pipeline}
\end{figure}

Figure~\ref{fig:compilation-pipeline} outlines the evaluation pipeline.
Initially, PolyBench C benchmarks were converted to MLIR using Polygeist,
with either \lstinline|SMALL| or \lstinline|MEDIUM| size configuration
based on the host machine.
Subsequently, the MLIR code was compiled into Wasm either via \tool
or through LLVM, applying standard MLIR optimizations.
We have enabled common subexpression elimination, loop
invariant code motion, loop invariant subset hoisting, control flow sinking,
and sparse conditional constant propagation.
In addition, some optimizations, such as constant folding, are applied by MLIR
by default. 
To fully leverage the optimizations available in LLVM,
we use \lstinline|-O3| optimization flag for LLVM compilation.
For both \tool and LLVM paths, we applied Binaryen optimizations 
with the default optimization level, \lstinline|-O2|.

We evaluate performance by executing the generated Wasm binaries on two
distinct hardware platforms. 
For testing on high-performance platforms, we use a MacBook M2 Pro running the
wasmtime~\cite{wasmtime} runtime in both interpreter and AoT compilation
modes with the \lstinline|MEDIUM| PolyBench dataset configuration. 
To measure performance on embedded systems, we deploy our experiments to an Apollo4 Blue Plus
(Arm Cortex-M4 with 2MB flash memory and 384KB RAM) running the
WAMR~\cite{wamr} runtime on Zephyr RTOS~\cite{zephyr}, using the
\lstinline|SMALL| PolyBench dataset due to memory constraints. 
In all cases,
we apply the highest available optimization levels for AoT compilation. 
We measure execution time as the interval between host function calls
surrounding each benchmark kernel, consistent with the measurement methodology
in the original PolyBench C implementation.

\subsection{Experiment Results}
\label{sec:experiment-results}

\begin{figure}[h]
  \begin{subfigure}[t]{\textwidth}
    \centering
    \includegraphics[width=\textwidth]{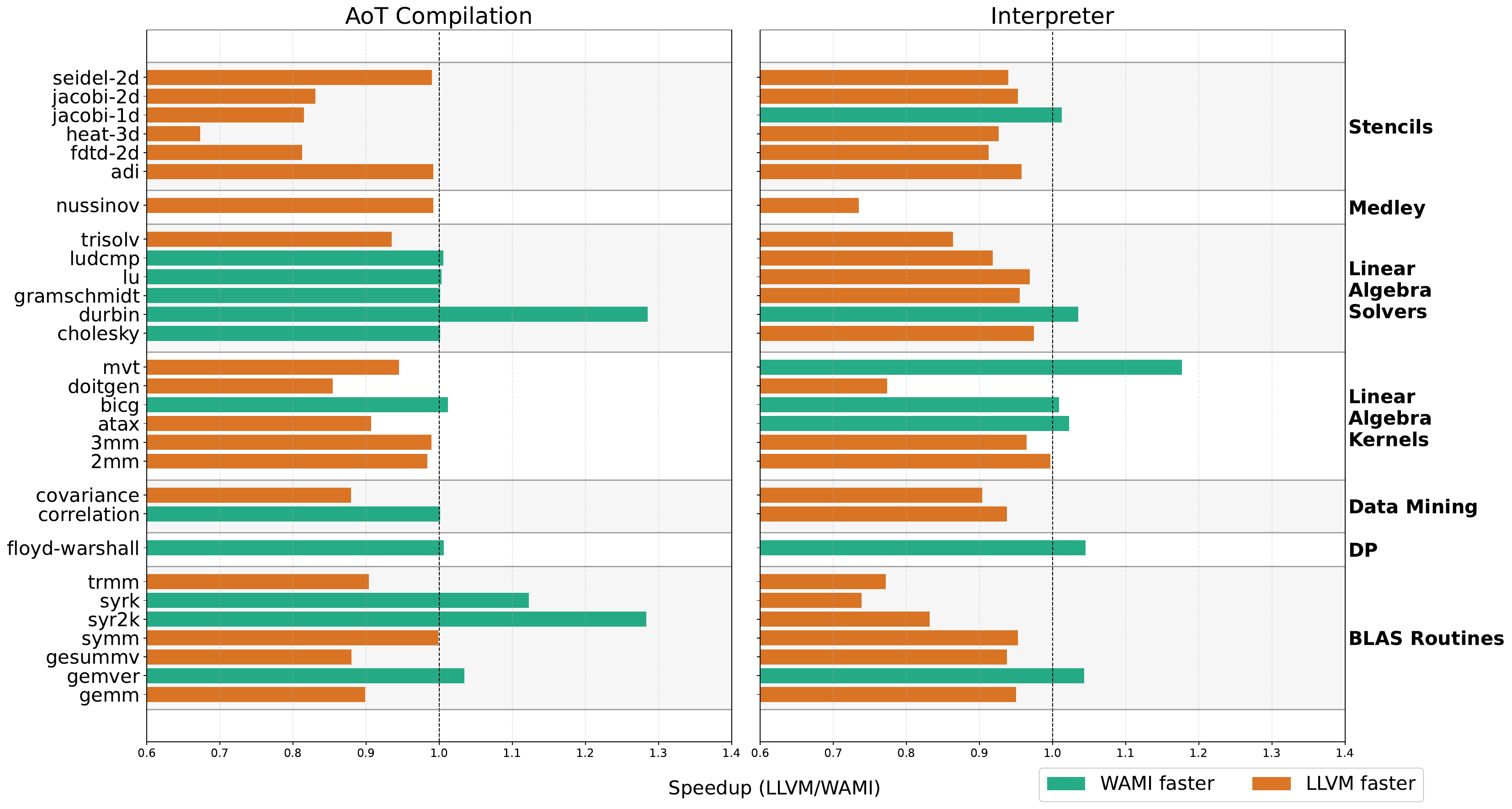}
    \caption{Wasmtime}
    \label{fig:wasmtime}
  \end{subfigure}
  \begin{subfigure}[t]{\textwidth}
    \centering
    \includegraphics[width=\textwidth]{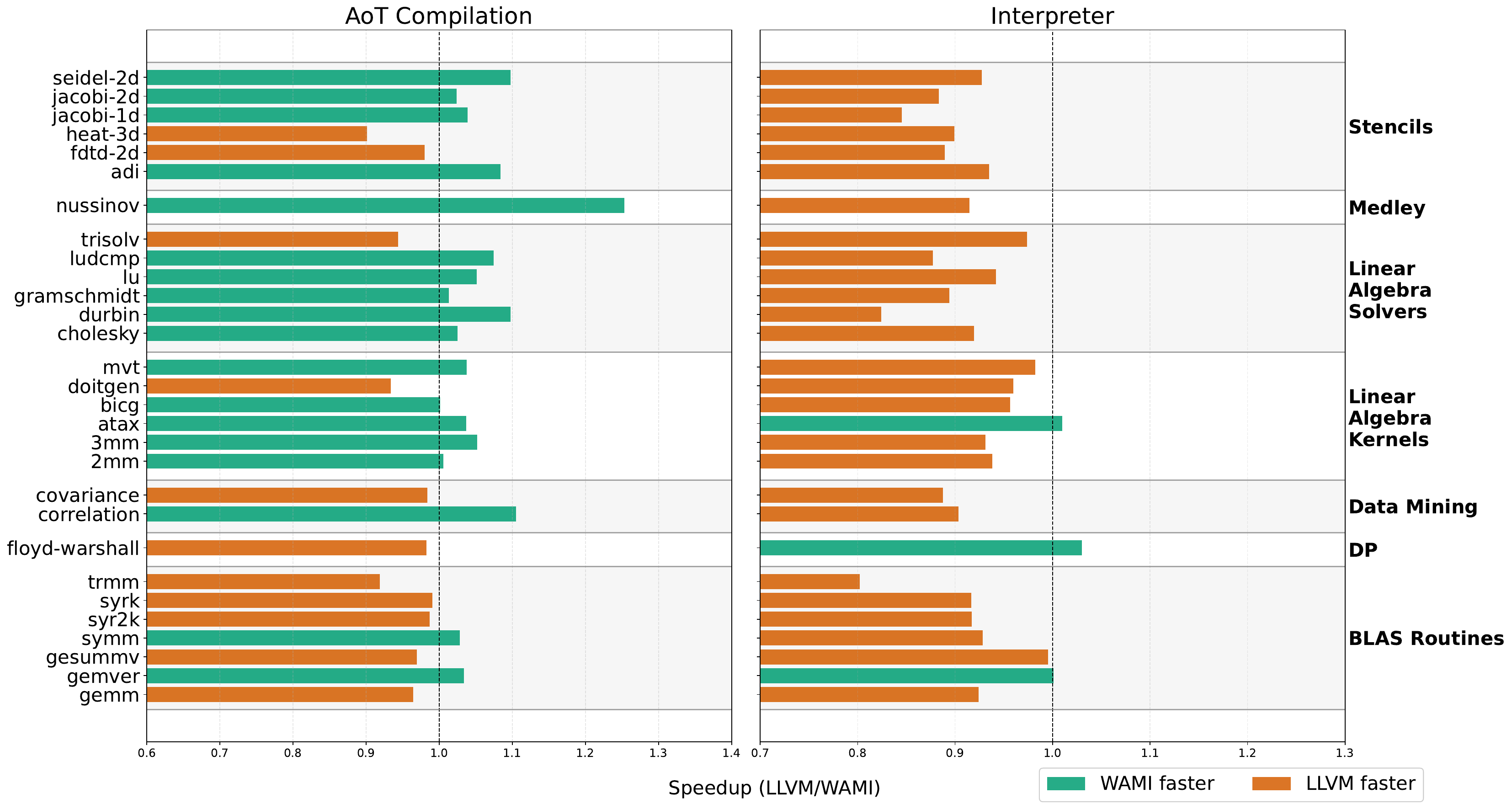}
    \caption{WAMR}
    \label{fig:wamr}
  \end{subfigure}
  \caption{Speedup of Wasm binaries produced by \tool over LLVM across different runtimes and configurations.}
  \label{fig:speedup}
\end{figure}

Figure~\ref{fig:speedup} illustrates the speedup of code compiled by \tool over
that compiled by LLVM.
Each bar shows the performance of the Wasm binary produced by \tool
normalized to the LLVM-generated code.
With interpreters,
\tool exhibits mean slowdown of 6.7\% and 7.7\% on Wasmtime and WAMR,
respectively.
With AoT compilation, it demonstrates mean slowdown of 4.1\% on Wasmtime and
1.9\% speedup on WAMR.

Overall, the performance of \tool-generated code is not significantly slower
than that of LLVM-generated code, being 7.7\% slower in the worst case,
despite \tool's lack of manually implemented optimization passes.
Thanks to existing optimizations in MLIR and Binaryen;
by modeling Wasm in MLIR, we inherit all general MLIR optimizations at no
additional implementation cost; 
similarly, being within the Wasm ecosystem grants us access to Binaryen
optimizations. 

On WAMR with AoT compilation, the performance of \tool-generated code is
competitive with LLVM, being 1.9\% faster.
The competitive performance stems from WAMR's AoT compiation
process. 
Since WAMR's AoT compiler 
leverages the LLVM backend to produce
optimized machine code from Wasm binaries, any sophisticated LLVM optimizations
missed by \tool can be still applied by the time of execution,
effectively neutralizing potential optimization differences
present in the original Wasm binaries.
To validate the hypothesis, we conducted the same experiment without AoT
compilation and observed the average slowdown by 17.0\% compared to LLVM-based
compiler, suggesting that the AoT optimizations contribute significantly to the
comparable performance.

On Wasmtime with AoT compilation as well as on interpreters, \tool-generated
code exhibits slightly lower performance compared to LLVM-generated code.
This is because Wasmtime's AoT compiler uses Cranelift~\cite{cranelift} as its
backend, applying less aggressive optimizations than LLVM
to achieve fast compilation times.
Additionally, interpreters execute Wasm binaries directly without optimizations.
Consequently, \tool-generated code is less competitive against LLVM-generated
code in Wasmtime environments than when AoT compilation is performed on WAMR.
Nevertheless, thanks to optimizations from MLIR and Binaryen,
it still maintains robust overall performance.

\begin{figure}[ht]
  \centering
\begin{minipage}[t]{0.5\textwidth}
  \begin{subfigure}[t]{\textwidth}
    \centering
    \begin{lstlisting}[style=mlir]
affine.for %i = 0 to 100 {
  // convert loop index to f64
  %0 = arith.index_cast %i : index to i32 
  %1 = arith.sitofp %0 : i32 to f64 
  ...
}
    \end{lstlisting}
    \caption{MLIR code snippet from the \lstinline|covariance| testcase}
    \label{fig:mlir-covariance}
  \end{subfigure}
  \vspace{.5em}

  \begin{subfigure}[t]{\textwidth}
    \centering
    \begin{lstlisting}[style=wasm]
(block $block 
  ... ;; initialize loop variables
  (loop $loop 
    ...  ;; load loop index
    local.get $i
    f64.convert_i32_s ;; cast to f64
    ... ;; branch back to the loop
  )
)
    \end{lstlisting}
    \caption{Wasm output by our compiler, without optimization.}
    \label{fig:wasm-compiler}
  \end{subfigure}
\end{minipage}
\hfill
\begin{minipage}[t]{0.45\textwidth}
  \begin{subfigure}[t]{\textwidth}
    \centering
    \begin{lstlisting}[style=wasm]
(block $block 
  ...
  (loop $loop
    ...  ;; load a variable that stores the f64-casted loop index local.get $temp
    local.get $temp
    f64.const 1.0
    f64.add ;; increment the variable by 1.0
    local.set $temp
    ... ;; branch back to the loop
  )
)
    \end{lstlisting}
    \caption{Wasm output by LLVM. Strength reduction optimization is applied.}
    \label{fig:wasm-llvm}
  \end{subfigure}
\end{minipage}

  \caption{Comparison of Wasm code generated by the Wasm dialect-based compiler and LLVM.}
  \label{fig:wasm-llvm-comparison}
\end{figure}

Moreover, applying LLVM optimizations may not always be optimal given that many
Wasm runtimes can eventually apply machine-specific optimizations through JIT or
AoT compilation.
As a portable bytecode format, Wasm executed in various environments,
some of which have different execution models than the traditional CPU
that LLVM optimizations are designed for.
In resource-constrained microcontrollers, for instance,
the classic stack-based interpreter of WAMR may be preferred.
It is slower than the default interpreter, which pre-processes the Wasm code
into a register-based format~\cite{wamr-fast-interpreter}, but consumes about
2\(\times\) less memory~\cite{wamr-interpreter-comparison}.
LLVM optimizations, originally designed for traditional CPU execution models,
may degrade performance in such environments due to the mismatch between 
their assumed execution model and the actual
runtime.

\begin{figure}[ht]
  \centering
  \includegraphics[width=\textwidth]{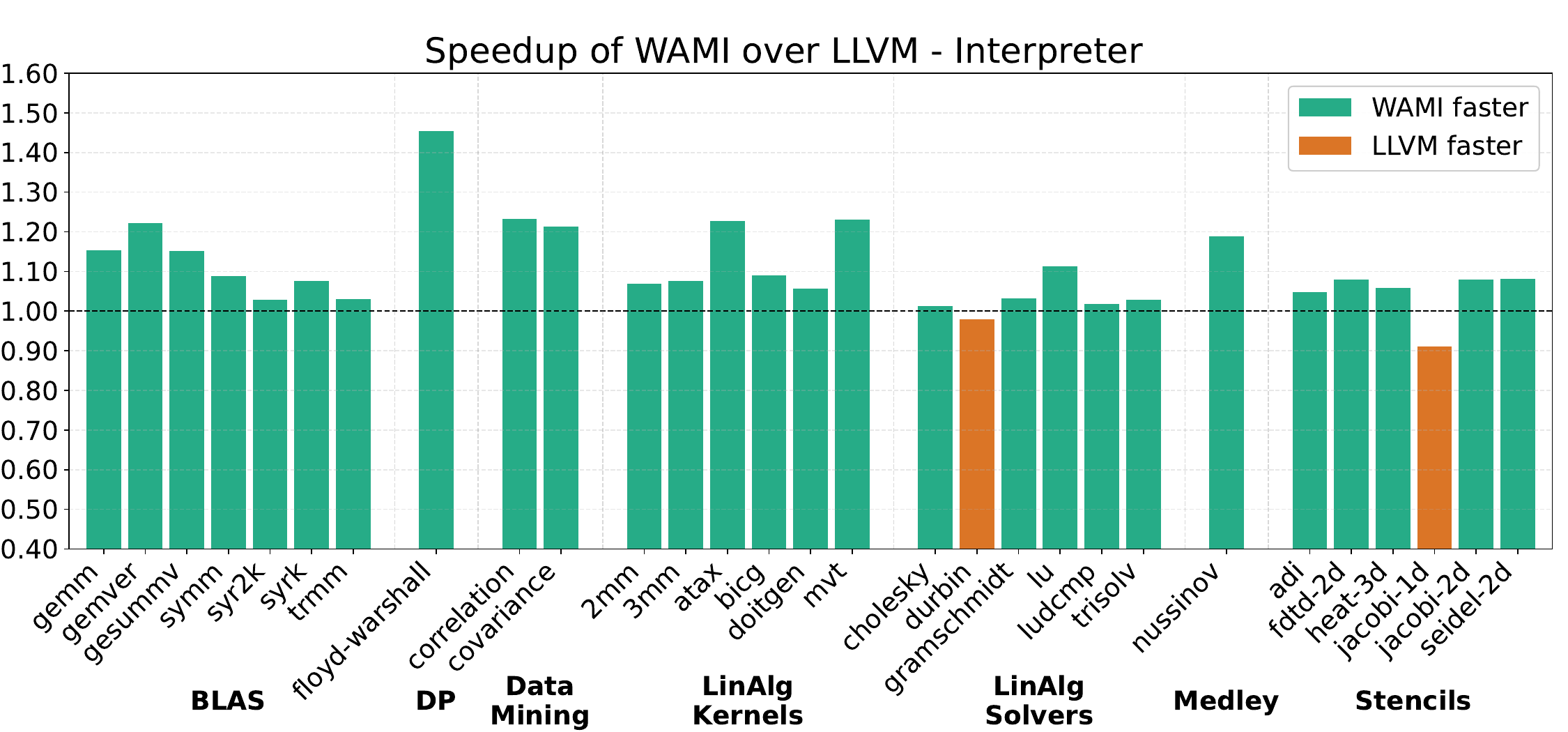}
  \caption{Speedup of \tool-generated code over LLVM on the WAMR classic interpreter.}
  \label{fig:wamr-classic-interpreter}
\end{figure}

Figure~\ref{fig:wasm-llvm-comparison} shows such an example.
The original MLIR code casts the current loop index to a floating-point
number and uses it throughout the loop.
\tool, without optimizations, simply retrieves the current loop
index (stored in the local variable \lstinline|$i|) and applies
\lstinline|f64.convert_i32_s| to convert it to a floating-point number.
LLVM applies strength reduction optimization~\cite{cooper2001operator}, replacing
the casting with a \lstinline|local.get| instruction, by introducing a new local
variable \lstinline|$temp| that increments by 1.0 in each iteration.
While this is a valid optimization in most traditional execution environments,
it slows down the performance of the stack-based interpreter, as newly
introduced local variable accesses are translated into memory accesses, which
are more expensive than the casting instruction in the original code. 
As a result, on the classic stack-based interpreter of WAMR,
\tool achieves mean speedup of 10.0\% over LLVM, as shown in
Figure~\ref{fig:wamr-classic-interpreter}. 

\subsection{Summary}

Our evaluation highlights that,
despite not implementing any optimization passes manually
and not leveraging LLVM's extensive optimization infrastructure,
\tool consistently generates code with reasonable performance.
This outcome primarily results from leveraging general optimization passes 
provided by MLIR, optimizations available through Binaryen,
and inherent properties of the Wasm execution model.
These findings suggest that extensive LLVM compiler optimizations, traditionally
seen as critical, are less significant when targeting Wasm from MLIR, indicating
that our Wasm dialect-based compilation pipeline is both practical and feasible
for incremental adoption without noticeable performance degradation.

\section{Discussion}
\label{sec:discussion}

The primary limitation of our work arises from the relatively immature ecosystem
for general-purpose languages within MLIR. 
Currently, MLIR is predominantly
used for domain-specific compilers, and existing benchmark suites primarily
represent domain-specific applications, such as polyhedral optimizations.
Consequently, there are a limited number of high-level dialects available that
could directly benefit from the pipeline proposed in this study.

Nevertheless, there is a growing community interest in broadening MLIR's
capabilities to support general-purpose programming. 
Efforts such as integrating
functional programming dialects into MLIR~\cite{lucke2021integrating}
and modeling JavaScript dialect for MLIR~\cite{jsir} highlight
this ongoing trend.
Moreover, although MLIR lacks general modeling for non-local control flow yet, 
MLIR documentation explicitly categorizes it as an operation behavior and
welcomes patches for this feature~\cite{mlir-side-effects},
demonstrating that such high-level programming constructs are within MLIR's
design scope.

Complementing these ongoing efforts, we believe that our approach to compiling
to Wasm with advanced language features, without compromising performance, will 
incentivize further integration of general-purpose languages within the MLIR
framework. 
Future work involves extending dialect support to include emerging language
proposals, integrating additional high-level language features into
MLIR dialects, and actively contributing to standardization initiatives aimed at
strengthening general-purpose programming capabilities across the MLIR
ecosystem.

\section{Related Work}
\label{sec:relatedwork}

\paragraphb{Compilation of High-Level Languages to Wasm}
A variety of compilers exist that compile high-level languages to Wasm.
Many languages, including C, C++, Rust, Zig, and Haskell, use LLVM to target Wasm.
They do not use high-level features of Wasm, such as Garbage Collection,
although there has been an attempt \cite{clangwebasembly}, which
is no longer maintained.
On the other hand, Kotlin~(Kotlin/Wasm)~\cite{kotlinwasm}, Dart~\cite{dartwasm},
OCaml~\cite{andres2023wasocaml}, Java~\cite{javawasm}, Go~\cite{go-wasm}, build
a custom compilation pipeline that compiles to Wasm instead of going through a
shared intermediate representation.
This makes it easier to support high-level Wasm features.
For example, most of them support garbage collection at least as an experimental
feature.
However, they lack the community support that shared compiler frameworks provide.
Languages such as AssemblyScript~\cite{assemblyscript} and Grain~\cite{grain}
compile to Binaryen IR and use Binaryen as their backend.
However, these compilers only support Wasm as their target.
Our work is different from them as it takes advantage of the benefits of shared
compiler frameworks but also is capable of generating Wasm code with high-level
features.

\paragraphb{Preserving High-Level Abstractions through MLIR}
The key advantage of MLIR is its ability to maintain high-level abstractions
across the entire compilation pipeline. Many compilers use MLIR to
preserve these abstractions, enabling optimizations that depend on an in-depth
semantic understanding of the program. 
For instance, recent research~\cite{pienaar2020mlir, zhu2021disc,
golin2024towards, jin2020compiling, bik2022compiler} utilizes MLIR to perform
tensor optimizations. 
Additionally, frameworks such as
Polygeist~\cite{moses2021polygeist} leverage MLIR for polyhedral optimizations.
While implementing Wasm-specific high-level optimizations with MLIR is the scope of this paper, it represents an intriguing direction for future
investigation.
 
Dialect-based design in MLIR not only supports optimization but also ensures the
preservation of high-level semantics through successive lowering stages, thus
avoiding the need for semantic reconstruction at lower levels such as LLVM IR.
For example, MLIR's built-in \lstinline|Vector| dialect~\cite{mlir-vector}
enables direct translation from higher-level dialects like \lstinline|Affine|,
facilitating efficient vector instruction generation without relying on LLVM's
vectorization passes.
Similarly, the \lstinline|GPU| dialect~\cite{mlir-gpu} preserves GPU-specific
abstractions throughout the compilation process, removing the need for
regenerating GPU code from LLVM IR. 
Our approach adopts a similar strategy, using specialized Wasm dialects to
maintain high-level Wasm semantics for straightforward and effective code
generation.

\paragraphb{Codegen in MLIR}
The key characteristic of our work is utilizing MLIR for Wasm code generation
without relying on the LLVM backend. 
Although MLIR currently lacks robust support for machine-code generation
algorithms, it is widely adopted for code generation targeting domain-specific
architectures, where traditional machine-code generation techniques are often
unnecessary~\cite{circt}. 
Our approach follows this idea, leveraging the simplicity of Wasm's machine
model, which does not require sophisticated code generation algorithms, to
produce efficient code directly from MLIR. 
Previous efforts have explored generating x86 machine code from MLIR primarily
to accelerate Just-In-Time (JIT) compilation~\cite{jit-x86-mlir}, as well as
migrating LLVM-based code generation into MLIR to mitigate the \emph{lost in
translation} problem~\cite{move-llvm-codegen-slides}. 

\section{Conclusion}
\label{sec:conclusion}

We presented a novel compilation pipeline for Wasm by introducing new MLIR
dialects to model Wasm with dialect conversion passes to 
generate Wasm from high-level MLIR dialects.
This approach preserves high-level abstractions throughout the compilation,
enabling effective utilization of advanced Wasm features.
%
Our experimental results demonstrate that our compiler produces code with
reasonable performance, comparable to that generated by LLVM, even without
employing optimization passes implemented in LLVM. 
We believe our work is a step towards creating more modular and extensible
compiler infrastructures for high-level languages targeting Wasm.

\bibliographystyle{ACM-Reference-Format}
\bibliography{reference}

\end{document}